\documentclass[aps,superscriptaddress,floatfix,amsmath,amssymb,prl,twocolumn]{revtex4-2}

\usepackage{graphicx}
\usepackage{xcolor}
\usepackage{xspace}
\usepackage[colorlinks=true,plainpages=false,linkcolor=blue,urlcolor=blue,citecolor=blue,pdfpagemode=UseNone,pdfstartview=FitBH]{hyperref}
\usepackage{amsmath} 

\newcommand{\cvs}{CsV$_{3}$Sb$_{5}$\xspace}
\newcommand{\fref}[1]{Fig.\,\ref{#1}}

\begin{document}

\title{Pressure-dependent Electronic Superlattice in the Kagome-Superconductor CsV$_3$Sb$_5$}

\author{F. Stier}
\altaffiliation{These authors contributed equally to this work}
\affiliation{Institut für Festkörper- und Materialphysik, Technische Universität Dresden, 01062 Dresden, Germany}

\author{A.-A.~Haghighirad}
\altaffiliation{These authors contributed equally to this work}
\affiliation{Institute for Quantum Materials and Technologies, Karlsruhe Institute of Technology, Kaiserstr. 12, D-76131 Karlsruhe, Germany}

\author{G.~Garbarino}
\affiliation{ESRF, The European Synchrotron, 71, avenue des Martyrs, CS 40220 F-38043 Grenoble Cedex 9}

\author{S.~Mishra} 
\affiliation{Institut für Festkörper- und Materialphysik, Technische Universität Dresden, 01062 Dresden, Germany}
\affiliation{W\"urzburg-Dresden Cluster of Excellence ct.qmat, Technische Universit\"at Dresden, 01062 Dresden, Germany}

\author{N.~Stilkerich}
\affiliation{Institut für Festkörper- und Materialphysik, Technische Universität Dresden, 01062 Dresden, Germany}
\affiliation{Max Planck Institute for Chemical Physics of Solids, Dresden, Germany}

\author{D.~Chen}
\affiliation{Max Planck Institute for Chemical Physics of Solids, Dresden, Germany}
\affiliation{College of Physics, Qingdao University, Qingdao 266071, China}

\author{C.~Shekhar}
\affiliation{Max Planck Institute for Chemical Physics of Solids, Dresden, Germany}

\author{T.~Lacmann}
\affiliation{Institute for Quantum Materials and Technologies, Karlsruhe Institute of Technology, Kaiserstr. 12, D-76131 Karlsruhe, Germany}

\author{C.~Felser}
\affiliation{Max Planck Institute for Chemical Physics of Solids, Dresden, Germany}
\affiliation{W\"urzburg-Dresden Cluster of Excellence ct.qmat, Technische Universit\"at Dresden, 01062 Dresden, Germany}

\author{T.~Ritschel}
\affiliation{Institut für Festkörper- und Materialphysik, Technische Universität Dresden, 01062 Dresden, Germany}

\author{J.~Geck}
\email{jochen.geck@tu-dresden.de}
\affiliation{Institut für Festkörper- und Materialphysik, Technische Universität Dresden, 01062 Dresden, Germany}
\affiliation{W\"urzburg-Dresden Cluster of Excellence ct.qmat, Technische Universit\"at Dresden, 01062 Dresden, Germany}

\author{M.~Le~Tacon}
\email{matthieu.letacon@kit.edu}
\affiliation{Institute for Quantum Materials and Technologies, Karlsruhe Institute of Technology, Kaiserstr. 12, D-76131 Karlsruhe, Germany}

\date{\today}

\begin{abstract}


We present a high-resolution single crystal x-ray diffraction study of kagome-superconductor \cvs, exploring its response to variations in pressure and temperature. We discover that at low temperatures, the structural modulations of the electronic superlattice, commonly associated with charge-density-wave order, undergo a transformation around $p \sim$ 0.7 GPa from the familiar $2\times2$ pattern to a long-range-ordered modulation at wavevector $q=(0, 3/8, 1/2)$. 
Our observations align with inferred changes in the CDW pattern from prior transport and nuclear-magnetic-resonance studies, providing new insights into these transitions. Interestingly, the pressure-induced variations in the electronic superlattice correlate with two peaks in the superconducting transition temperature as pressure changes, hinting that fluctuations within the electronic superlattice could be key to stabilizing superconductivity. 
However, our findings contrast with the minimal pressure dependency anticipated by ab initio calculations of the electronic structure. They also challenge prevailing scenarios based on a Peierls-like nesting mechanism involving van Hove singularities.

\end{abstract}

\maketitle

Understanding the correlation between unconventional superconductivity (SC) and spatial electronic modulations, encompassing magnetic, nematic, or charge-density-wave (CDW) orders, represents one of the key challenges in contemporary condensed matter research~\cite{Fernandes_ARCMP2019, Agterberg_2020, Neupert_NatPhys2022}. Consequently, extensive endeavors are underway to unravel the underlying physics in quantum materials like cuprates~\cite{Keimer_Nature2015}, nickelates~\cite{Tam_Natmat2022,Krieger_PRL2022}, iron or nickel pnictides~\cite{Yi_PRL2018, Souliou_PRL2022, Lacmann_PRB2023}, transition metal dichalcogenides~\cite{Leroux_PRB2015} or heavy fermion compounds~\cite{Aishwarya_Nature2023}, where the relation between SC and CDW phenomena can be examined in different settings.

In this framework, a new and extremely intriguing class of materials has recently been discovered: the layered kagome compounds \textit{A}V$_3$Sb$_5$ (\textit{A}=K, Cs, Rb)~\cite{Ortiz_PRL2020}. Characterized by an electronic structure featuring flat bands, van Hove singularities (vHs), Dirac cones, and non-trivial band topology, these materials offer a unique platform for exploring novel electronic states of matter with intertwined orders~\cite{Neupert_NatPhys2022}. Particularly noteworthy is the detection of an electronic superlattice below 94 K,~\cite{Ortiz_PRL2020} which is commonly associated with a CDW, and that is succeeded by superconductivity below $T_{c} \simeq 2.7$~K in \cvs -- a revelation that has sparked considerable excitement.
This excitement has been further substantiated by recent accounts of the absence of Kohn anomalies in the phonon dispersion associated with the formation of the CDW~\cite{Li_PRX2021, Subires_NatCom2023}, chirality~\cite{Guo_Nature2022}, time-reversal symmetry breaking or electronic nematicity~\cite{Nie_Nature2022} of the electronic superlattice. 
Consequently, while we refer to the electronic modulation in \cvs as a CDW in the following, we recognize that it is often rightly considered unconventional and that the underlying electronic order may be substantially more complex. As a matter of fact, numerous important aspects of this CDW and of its relationship with SC in \cvs remain elusive and subject to intense debate~\cite{Nie_Nature2022, Zheng_Nature2022, Asaba_2023,Frachet_2023, Liu_2023}. The origin of the CDW and the nature of the interplay between electronic and lattice degrees of freedom in these materials therefore requires further elucidation~\cite{Li_PRX2021, Zhong_NatCom2023, Subires_NatCom2023, Ritz_PRB2023}. 

\begin{figure*}[t!]
  \includegraphics[width=0.95\textwidth]{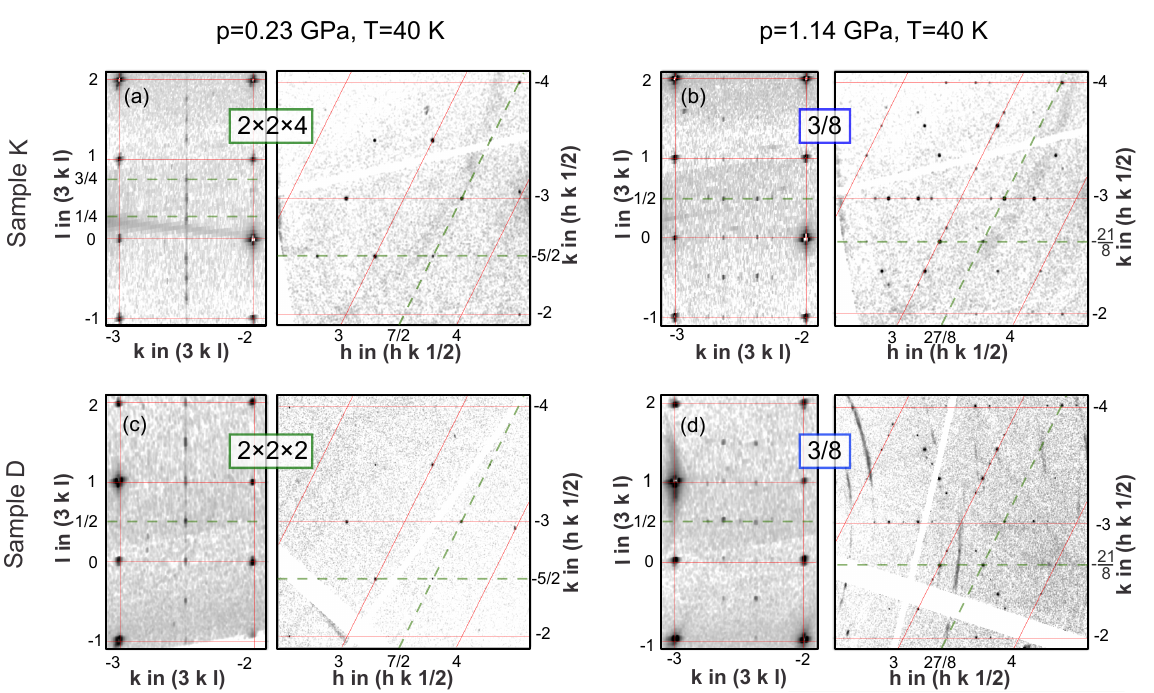}
\caption{Temperature and pressure dependence of the superlattice modulation observed for sample $K$ (top) and $D$ (bottom). $hl$- and the $hk$-cuts through XRD intensity distributions in the reciprocal space are shown.
At 40~K and $p=0.23$~GPa, the two samples both display a $2\times2$ modulation along $ab$ but a different superstructure along $c$.  Increasing the pressure at 40~K above 0.7 GPa, here shown for 1.14 GPa, induces a new 3/8-CDW-order, which is the same for both samples.
The green dashed lines in panel (a) indicate the $2\times2$x4 peaks, in panel (b) the $2\times2$x2 peaks, and for (c) and (d) the 3/8-peaks.} 
  \label{fig1}
  \vspace{-0.5cm}
\end{figure*}

To address such issues, pressure tuning has proven to be a valuable tool in quantum materials research~\cite{Souliou_2020,Pimenta_APR2023,Mao_RPM2018}, providing a reversible method to adjust the balance between competing energy scales and enabling the investigation of complex phase diagrams in these materials without introducing chemical disorder. This approach is without doubt most relevant in the case of kagome superconductors, since both SC and CDW transition temperatures have been reported to strongly depend on hydrostatic pressure, displaying intriguing behavior already for relatively modest pressures~\cite{Yu_NatCom2021, Feng_NPJQM2023, Zheng_Nature2022}. This is demonstrated across a diverse array of measurements encompassing electrical- and magnetotransport, magnetic susceptibility, nuclear magnetic and quandrupole resonances (NMR and NQR), x-ray diffraction (XRD), and muon spin resonance~\cite{Yu_NatCom2021, Feng_NPJQM2023, Zheng_Nature2022,Mielke_Nat2022,Guguchia_NPJ2023} which have revealed that both the superconducting transition temperature ($T_c$) and the upper critical field $H_{c2}$ as a function of pressure $p$ exhibit a distinctive double-dome feature below $p<2.0$\,GPa , characterized by a first maximum of $\sim$ 7 K at $p_1 \simeq 0.7$ GPa, succeeded by a second maximum at $p_2\simeq 2$ GPa with $T_c\simeq$ 8~K~\cite{Zheng_Nature2022, Feng_NPJQM2023, Chen_PRL2021, Du_PRB2021}. The CDW transition temperature appears monotonically suppressed with pressure, but from a marked reduction of the magnetoresistance \cite{Yu_NatCom2021} and the appearance of new NMR lines above $p_1$, changes in the electronic modulation pattern have been indirectly inferred.

Nonetheless, a thorough examination of the pressure dependence of CDW superstructures across various temperatures -- essential for clarifying the microscopic physics at play -- has yet to be undertaken. Here we close this gap, by reporting a detailed pressure- and temperature-dependent single crystal x-ray diffraction (XRD) study of the crystal lattice and its superstructures in \cvs. 
We report the gradual suppression of the familiar $2\times2$ CDW modulation with pressure, alongside the appearance of a new type of modulation with a completely different pattern and periodicity above $p_1$. The two orders coexist for a while before the disappearance of the former. At 25 K, no CDW survives above $\sim$ 1.7 GPa. 

The clear change of the CDW pattern we report in the investigated pressure range, however, contrasts with the modest impact of pressure on the electronic structure as computed -- using the experimentally determined pressure-dependent lattice parameters -- by first-principles calculations. In particular, the van Hove singularities below the Fermi energy show minimal pressure dependence in the investigated range. This discrepancy challenges weak coupling scenarios that attribute the formation of electronic superlattices in kagome metals \textit{A}V$_3$Sb$_5$ (\textit{A}=K, Cs, Rb) to a Peierls-like nesting of van Hove singularities~\cite{Tan_PRL2021,Kang_NaturePhysics2022}.

\begin{figure}[t!]
 \includegraphics[width=\linewidth]{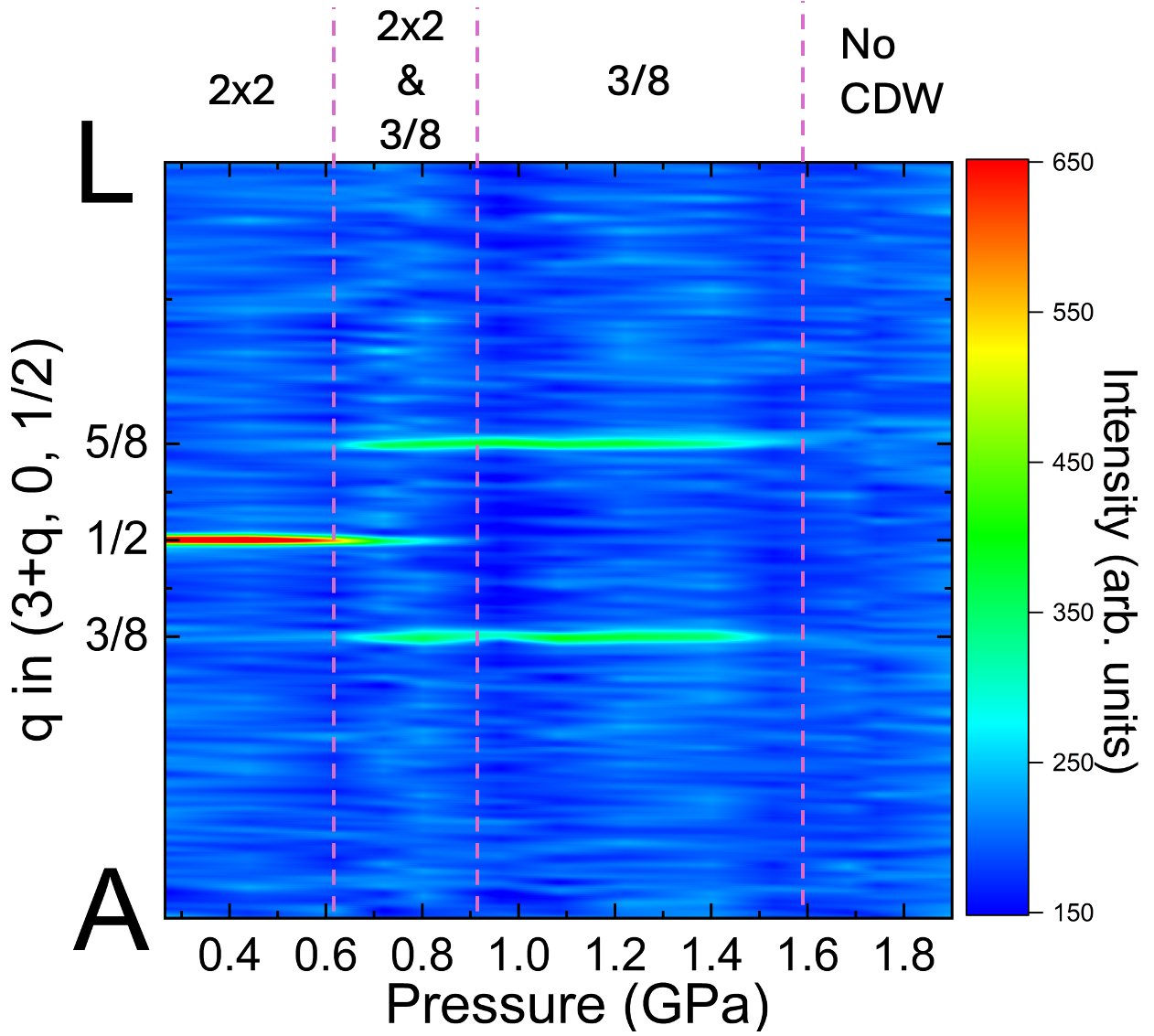}
  \caption{Pressure dependence of the CDW order in \cvs. The contour plot shows the XRD-intensity along the $A-L$ reciprocal direction of the hexagonal cell (obtained from the averaging of the intensity measured at 25\,K along the six reciprocal $A-L$ directions from the (3, 0, 0.5) A point). The intensity at the (3.5, 0, 0.5) point (L-point of the hexagonal cell) below 0.7\,GPa signals the $2\times2\times2$ order, which transforms into the 3/8 modulation with increasing pressure revealed by the intensity at $q=3/8$ and $q=5/8$ above 0.7\,GPa.}
  \label{fig3}
   \vspace{-0.5cm}
\end{figure}

We investigated single crystals from two distinct batches, referred to as batch $D$ and batch $K$ in the following, synthesized and characterized, as described in the supplementary information~\cite{SOM}. 
High-resolution XRD measurements as a function of $T$ and $p$ were performed at the ID15B beamline of the ESRF in Grenoble, France, using a monochromatic x-ray beam with an energy of 30.17 keV, 
focused down to 2x4 $\mu$m$^2$ at the sample position. The diffraction patterns were measured with an EIGER2 X 9M CdTe flat panel detector. Diamond anvil cells (DACs) with helium as a pressure transmitting medium were used to apply pressures of up to 27~GPa under close to hydrostatic conditions. 
In each DAC, one sample of batch $D$ and one of batch $K$ were loaded into the sample space, so that both samples were in exactly the same experimental condition during our pressure and temperature dependent XRD measurements. Further details are given in the supplementary information~\cite{SOM}.

At room temperature and ambient pressure, both samples exhibit a hexagonal $P6/mmm$ structure, in full agreement with previously published results~\cite{Ortiz_PRM2019}. In the following, reflections will be always indexed referring to this structure, even though there are structural transitions as a function of $p$ and $T$~\cite{SOM}. In agreement with previous reports~\cite{Ortiz_PRL2020}, both samples develop a $2\times2$ superstructure within the $ab$-planes when cooled below 94~K at ambient or small pressure. 
This CDW transition is signalled by additional satellite reflections, as demonstrated in panels (a) and (c) of \fref{fig1}. Interestingly, as can also be observed in these panels, the structural modulation of the two set of samples differs along the $c$-direction. At 40~K and $p=0.23$\,GPa, despite special care taken to ensure they are precisely in the same experimental condition and sharing the same $pT$-history, sample $K$ exhibits a $2\times2\times4$ modulation, whereas sample $D$ displays a $2\times2\times2$ order. This observation supports earlier conclusions that the orders along $c$ are metastable~\cite{Xiao_PRR2023}, while the $2\times2$ modulation within the $ab$-layers is a defining feature of the electronic ordering instability of \cvs at ambient pressure~\cite{Ortiz_PRX2021}.
In both cases, the superstructure reflections are best indexed as single-Q modulations of an orthorhombic $Cmmm$ phase~\cite{SOM}, of symmetry lower than the original $P6/mmm$ in agreement with findings of Ref.~\onlinecite{Stahl_PRB22}.

The major experimental finding of our study is represented by the data in the right panels (b) and (d) of Fig.\,\ref{fig1}: Irrespective of the periodicity of the modulation along the $c$-axis, an increase of the pressure $p$ to 0.7~GPa at constant T=40~K induces in both samples a transition into a new phase (see also Supplementary Fig. 8~\cite{SOM}) characterized by the \textit{same} commensurate modulation vector $q=(0, \frac{3}{8}, \frac{1}{2})$. We will refer to this as the '3/8' phase in the following. 
This observation aligns very well with recent findings from $^{51}V$ NMR~\cite{Zheng_Nature2022} and $^{121/123}Sb$ NQR~\cite{Feng_NPJQM2023}, which reported in this very pressure range the appearance of new spectral features that are distinct from those of the $2\times2$ modulation and have been interpreted as evidence of a new type of CDW order. Although these local probes cannot directly determine the ordering wave vector, proposals for an incommensurate structure~\cite{Feng_NPJQM2023} or a unidirectional order with a period of $4\times a$~\cite{Zheng_Nature2022} have been suggested. While our XRD data confirms a pressure-driven change in the electronic order, we unambiguously reveal here that the $p$-induced ordering is commensurate and different from the suggested $4\times a$. 

With a further increase in $p$ at constant $T$, the intensities of the superlattice peaks of the 3/8-phase gradually diminish and vanish above 1.7~GPa, again in good agreement with the nuclear resonance data~\cite{Zheng_Nature2022, Feng_NPJQM2023}. According to our structural refinement, the 3/8 phase is accompanied with a small monoclinic distortion~\cite{SOM}.

To further elucidate the evolution of the CDW order with respect $p$, we show in \fref{fig3} the pressure dependence of the XRD-intensity along the high symmetry $A-L$ line of the hexagonal cell (starting from the reciprocal \textit{A} point at (3, 0, 0.5)) as a function of $p$ at constant $T=25$~K.
With $p$ increasing towards 0.7~GPa, 
the satellite reflections of the $2\times2$ in-plane order are gradually suppressed. As can be observed in \fref{fig3}
, the intensity of this peak weakens significantly around 0.7\,GPa, but remains detectable up to 0.9\,GPa. At the same time, the new 3/8 modulation emerges around 0.7\,GPa, i.e., the $2\times2$ and the 3/8 phases coexist between 0.7 and 0.9\,GPa. Beyond 0.9\,GPa, the $2\times2$ disappears completely, and by further increasing $p$ also the scattering due to the 3/8 modulation is progressively reduced until it falls below our detection limit at 1.7~GPa. 
The data shown in \fref{fig3} further reveals that, within the error of our experiment the positions of the satellite peaks corresponding to the $2\times2$ and the 3/8 order do not change with varying $p$ and that, despite the strong suppression of the intensity, the satellite reflections always remain as sharp as the Bragg reflections and are essentially resolution limited~\cite{SOM}.
This is true both for in- and out-of-plane directions.

\begin{figure}[b!]
  \includegraphics[width=\linewidth]{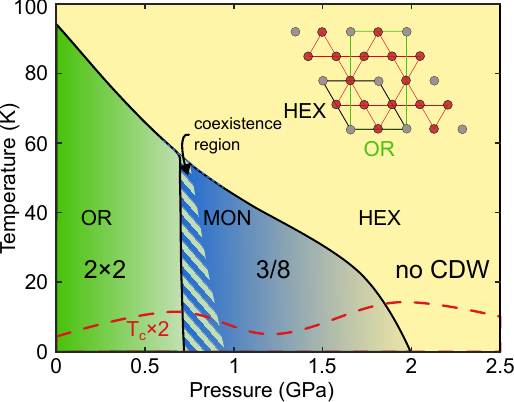}
  \caption{Electronic phase diagram of \cvs in the low-pressure and low-temperature region. The orthorhombic $2\times2\times2$ and $2\times2\times4$ phases, summarized as $2\times2$, exist in the green region below $\approx0.7$~GPa. The new monoclinic phase with the 3/8-order is indicated in blue. Both phases coexist in a narrow pressure range as indicated by the striped area. 
  The slope at low pressure ($dT_{CDW}/dp \sim -75K/GPa$) follows that dertermined from thermodynamic measurements~\cite{Frachet_2023}.
  The red dashed line sketches the evolution of the superconducting $T_c$ as a function of pressure (after Ref.~\onlinecite{Zheng_Nature2022}). \textit{Inset}: shows the hexagonal vs. orthorhombic lattice in the kagome plane.}
  \label{fig2}
   \vspace{-0.5cm}
\end{figure}

The results of our XRD studies below 100~K and up to 2.5~GPa are summarized in the $pT$-phase diagram presented in \fref{fig2}, where the stability regions of the $2\times2$ and the 3/8 orders are indicated by the green and blue areas, respectively. Comparing the pressure dependence of the electronic order to the pressure dependence of the superconducting $T_c$~\cite{Chen_PRL21, Feng_PRB21} reveals an interesting correlation: The pressure $p_1\simeq0.7$~GPa, where $T_c(p)$ shows its first maximum, aligns perfectly to the critical pressure where the $2\times2$ order is suppressed and the 3/8 order appears. Furthermore, the pressure $p_2\simeq2$~GPa of the second maximum of $T_c(p)$ coincides well with that at which the 3/8 phase disappears.  

The observation that every time an electronic order is suppressed with increasing $p$, the superconducting transition temperature $T_c(p)$ exhibits a maximum is remarkable. One obvious explanation to be considered is, of course, the competition between CDW and SC. Direct experimental evidence for this competition, for example in the form of a suppression of the CDW in the superconducting state is still lacking.
Nevertheless, the observation of a peak in $T_c(p)$ near the critical point of a distinct electronically ordered phase (here the $2\times2$ or the 3/8 phase) suggests that the corresponding fluctuations may contribute to supporting superconductivity. This conjecture certainly warrants careful examination in future investigations.

It is important to note that the consistently resolution-limited satellite reflections (cf. \fref{fig3}) do not necessarily imply the absence of fluctuations. The resolution-limited width of these reflections simply indicates that the coherent contribution to a given satellite reflection originates from a volume of about $10^4$ unit cells~\cite{SOM}. However, it remains indeterminate whether this volume constitutes a single contiguous entity or a series of smaller, distinct volumes maintaining a fixed phase relation (see Ref. \cite{Hossain:2013a}). In essence, despite the sharpness of the satellite peaks, regions experiencing fluctuations may still exist. These fluctuations would manifest as an exceedingly weak diffuse scattering signal, beyond the detection capabilities of the current experimental setup.

The comparison of the present $p$-dependent XRD results to those of a recent doping dependent study~\cite{Kautzsch_npjQM23} is also very interesting:  In the doping dependent study, it was observed that replacing Sb with Sn destabilizes the $2\times2$ superlattice as well, leading to the emergence of a new superlattice at low temperatures with a modulation wave vector similar to that of the 3/8 phase observed here. However, there are notable distinctions between the doping and $p$-induced orders: The doping-induced phase exhibits short and highly anisotropic correlation lengths within the $ab$-layers that have been interpreted as signs of electronic nematicity. In contrast, the the resolution-limited superlattice modulations observed in pristine \cvs, including the pressure-induced 3/8 phase, did not reveal any detectable in-plane anisotropy.
%

In principle, Sn-substitution and $p$ do not represent equivalent modifications of \cvs. On a qualitative level, however, they both affect the $c/a$ ratio, which has been found to play a critical role for controlling the electronic properties~\cite{Frachet_2023, Ritz_PRB2023}. 
Due to the anisotropic layered structure of \cvs, hydrostatic pressure has indeed a pronounced effect on the c/a ratio, which can be determined precisely by our measurements. We find that a pressure as modest as 1\,GPa is sufficient to reduce c/a by about 3\% (cf. Supplementary Fig. 4~\cite{SOM}).


In good agreement with DFT calculations~\cite{Ritz_PRB2023}, our structural refinement reveals that the distance between apical Sb and the kagome layers is barely affected by $p=1$\,GPa, in stark contrast to the inter unit-cell distance between apical Sb-sites, which decreases by  about 5\%. These structural changes primarily affect the Sb 5$p_z$-states, as clearly demonstrated in the DFT band structures shown in Supplementary Fig. 11~\cite{SOM}
, which are based on our structural data. As can be observed in this figure, the bottom of the $\Gamma$-centered electron pocket and the top of the hole-like band at the A-point move significantly closer to the Fermi level with increasing $p$. However, the vHs at the M-point, which are below the Fermi-level, remain essentially unaffected. This behaviour is in very good agreement with previous theoretical studies~\cite{Wenzel:2023a, Ritz_PRB2023,Tsirlin_SciPost22}, where similar trends have been reported.
%

The drastic pressure dependence of the electronic superlattice reported here therefore implies that the nesting of vHs cannot be the primary stabilization mechanism in \cvs. If this was the case, the $2\times2$ order would be expected to remain entirely stable below 1\,GPa, at odds with our experimental observation. This conclusion is further supported by a detailed analysis of the Lindhard function from DFT, which shows 
that the electronic system alone does therefore not favor a specific ordering wave vector~\cite{Kaboudvand:2022aa}.
Even though the new 3/8 phase  has not been predicted by any first principles calculations to the best of our knowledge, it would be essential to check within DFT whether it could become more stable than a $2\times2$ CDW under pressure. Strong coupling scenarios involving strongly momentum-dependent electron-phonon coupling~\cite{Xie:2022a,Kaboudvand:2022aa,Ye:2022a} or lattice anharmonicity effects~\cite{gutierrez_2023} have been considered as a driving force behind the stabilization of a specific charge order instabilities in these materials. Along these lines, further advanced first-principles calculations of the temperature and pressure dependence of the phonon spectrum and of the electron-phonon coupling might be required to clarify the mechanism behind the dramatic pressure-induced change of modulation reported here.

In conclusion, the present study yields two major findings namely the correlation between the critical pressures of the electronic modulations and superconductivity in CsV$_3$Sb$_5$ and a transition from the $2\times2$ to a long-range charge ordered 3/8 phase at pressures below 1 GPa. Furthermore, our findings reveal pressure-induced structural modulations that appear to coincide with the trend in the superconductivity manifested in other experiments. These observations strongly substantiate the notion that critical fluctuations and complex electron-lattice interactions, extending well beyond conventional scenarios and models, might play a prominent role in shaping the phase diagram of \textit{A}V$_3$Sb$_5$.

\section*{Acknowledgements}

We acknowledge fruitful discussions with R. Fernandes, R. Heid, M. Merz and S. M. Souliou. This research received support from the Deutsche Forschungsgemeinschaft (DFG, German Research Foundation), under projects 422213477 (TRR 288 project B03) and 247310070 (CRC 1143 project C06 and B05). T.R., F.S., C.F., and J.G. express gratitude to the W\"urzburg-Dresden Cluster of Excellence on Complexity and Topology in Quantum Matter (ct.qmat EXC-2147, Project No. 390858490) for financial backing. Additionally, we extend our appreciation to the European Synchrotron Radiation Facility (ESRF) for providing access to their synchrotron radiation facilities.


\begin{thebibliography}{55}%
\makeatletter
\providecommand \@ifxundefined [1]{%
 \@ifx{#1\undefined}
}%
\providecommand \@ifnum [1]{%
 \ifnum #1\expandafter \@firstoftwo
 \else \expandafter \@secondoftwo
 \fi
}%
\providecommand \@ifx [1]{%
 \ifx #1\expandafter \@firstoftwo
 \else \expandafter \@secondoftwo
 \fi
}%
\providecommand \natexlab [1]{#1}%
\providecommand \enquote  [1]{``#1''}%
\providecommand \bibnamefont  [1]{#1}%
\providecommand \bibfnamefont [1]{#1}%
\providecommand \citenamefont [1]{#1}%
\providecommand \href@noop [0]{\@secondoftwo}%
\providecommand \href [0]{\begingroup \@sanitize@url \@href}%
\providecommand \@href[1]{\@@startlink{#1}\@@href}%
\providecommand \@@href[1]{\endgroup#1\@@endlink}%
\providecommand \@sanitize@url [0]{\catcode `\\12\catcode `\$12\catcode `\&12\catcode `\#12\catcode `\^12\catcode `\_12\catcode `\%12\relax}%
\providecommand \@@startlink[1]{}%
\providecommand \@@endlink[0]{}%
\providecommand \url  [0]{\begingroup\@sanitize@url \@url }%
\providecommand \@url [1]{\endgroup\@href {#1}{\urlprefix }}%
\providecommand \urlprefix  [0]{URL }%
\providecommand \Eprint [0]{\href }%
\providecommand \doibase [0]{https://doi.org/}%
\providecommand \selectlanguage [0]{\@gobble}%
\providecommand \bibinfo  [0]{\@secondoftwo}%
\providecommand \bibfield  [0]{\@secondoftwo}%
\providecommand \translation [1]{[#1]}%
\providecommand \BibitemOpen [0]{}%
\providecommand \bibitemStop [0]{}%
\providecommand \bibitemNoStop [0]{.\EOS\space}%
\providecommand \EOS [0]{\spacefactor3000\relax}%
\providecommand \BibitemShut  [1]{\csname bibitem#1\endcsname}%
\let\auto@bib@innerbib\@empty
\bibitem [{\citenamefont {Fernandes}\ \emph {et~al.}(2019)\citenamefont {Fernandes}, \citenamefont {Orth},\ and\ \citenamefont {Schmalian}}]{Fernandes_ARCMP2019}%
  \BibitemOpen
  \bibfield  {author} {\bibinfo {author} {\bibfnamefont {R.~M.}\ \bibnamefont {Fernandes}}, \bibinfo {author} {\bibfnamefont {P.~P.}\ \bibnamefont {Orth}},\ and\ \bibinfo {author} {\bibfnamefont {J.}~\bibnamefont {Schmalian}},\ }\bibfield  {title} {\bibinfo {title} {Intertwined vestigial order in quantum materials: Nematicity and beyond},\ }\href {https://doi.org/10.1146/annurev-conmatphys-031218-013200} {\bibfield  {journal} {\bibinfo  {journal} {Annual Review of Condensed Matter Physics}\ }\textbf {\bibinfo {volume} {10}},\ \bibinfo {pages} {133} (\bibinfo {year} {2019})}\BibitemShut {NoStop}%
\bibitem [{\citenamefont {Agterberg}\ \emph {et~al.}(2020)\citenamefont {Agterberg}, \citenamefont {Davis}, \citenamefont {Edkins}, \citenamefont {Fradkin}, \citenamefont {Van~Harlingen}, \citenamefont {Kivelson}, \citenamefont {Lee}, \citenamefont {Radzihovsky}, \citenamefont {Tranquada},\ and\ \citenamefont {Wang}}]{Agterberg_2020}%
  \BibitemOpen
  \bibfield  {author} {\bibinfo {author} {\bibfnamefont {D.~F.}\ \bibnamefont {Agterberg}}, \bibinfo {author} {\bibfnamefont {J.~C.~S.}\ \bibnamefont {Davis}}, \bibinfo {author} {\bibfnamefont {S.~D.}\ \bibnamefont {Edkins}}, \bibinfo {author} {\bibfnamefont {E.}~\bibnamefont {Fradkin}}, \bibinfo {author} {\bibfnamefont {D.~J.}\ \bibnamefont {Van~Harlingen}}, \bibinfo {author} {\bibfnamefont {S.~A.}\ \bibnamefont {Kivelson}}, \bibinfo {author} {\bibfnamefont {P.~A.}\ \bibnamefont {Lee}}, \bibinfo {author} {\bibfnamefont {L.}~\bibnamefont {Radzihovsky}}, \bibinfo {author} {\bibfnamefont {J.~M.}\ \bibnamefont {Tranquada}},\ and\ \bibinfo {author} {\bibfnamefont {Y.}~\bibnamefont {Wang}},\ }\bibfield  {title} {\bibinfo {title} {The physics of pair-density waves: Cuprate superconductors and beyond},\ }\href {https://doi.org/10.1146/annurev-conmatphys-031119-050711} {\bibfield  {journal} {\bibinfo  {journal} {Annual Review of Condensed Matter Physics}\ }\textbf {\bibinfo {volume} {11}},\ \bibinfo {pages} {231}
  (\bibinfo {year} {2020})}\BibitemShut {NoStop}%
\bibitem [{\citenamefont {Neupert}\ \emph {et~al.}(2022)\citenamefont {Neupert}, \citenamefont {Denner}, \citenamefont {Yin}, \citenamefont {Thomale},\ and\ \citenamefont {Hasan}}]{Neupert_NatPhys2022}%
  \BibitemOpen
  \bibfield  {author} {\bibinfo {author} {\bibfnamefont {T.}~\bibnamefont {Neupert}}, \bibinfo {author} {\bibfnamefont {M.~M.}\ \bibnamefont {Denner}}, \bibinfo {author} {\bibfnamefont {J.-X.}\ \bibnamefont {Yin}}, \bibinfo {author} {\bibfnamefont {R.}~\bibnamefont {Thomale}},\ and\ \bibinfo {author} {\bibfnamefont {M.~Z.}\ \bibnamefont {Hasan}},\ }\bibfield  {title} {\bibinfo {title} {Charge order and superconductivity in kagome materials},\ }\href {https://doi.org/10.1038/s41567-021-01404-y} {\bibfield  {journal} {\bibinfo  {journal} {Nature Physics}\ }\textbf {\bibinfo {volume} {18}},\ \bibinfo {pages} {137} (\bibinfo {year} {2022})}\BibitemShut {NoStop}%
\bibitem [{\citenamefont {Keimer}\ \emph {et~al.}(2015)\citenamefont {Keimer}, \citenamefont {Kivelson}, \citenamefont {Norman}, \citenamefont {Uchida},\ and\ \citenamefont {Zaanen}}]{Keimer_Nature2015}%
  \BibitemOpen
  \bibfield  {author} {\bibinfo {author} {\bibfnamefont {B.}~\bibnamefont {Keimer}}, \bibinfo {author} {\bibfnamefont {S.~A.}\ \bibnamefont {Kivelson}}, \bibinfo {author} {\bibfnamefont {M.~R.}\ \bibnamefont {Norman}}, \bibinfo {author} {\bibfnamefont {S.}~\bibnamefont {Uchida}},\ and\ \bibinfo {author} {\bibfnamefont {J.}~\bibnamefont {Zaanen}},\ }\bibfield  {title} {\bibinfo {title} {From quantum matter to high-temperature superconductivity in copper oxides},\ }\href {https://doi.org/10.1038/nature14165} {\bibfield  {journal} {\bibinfo  {journal} {Nature}\ }\textbf {\bibinfo {volume} {518}},\ \bibinfo {pages} {179} (\bibinfo {year} {2015})}\BibitemShut {NoStop}%
\bibitem [{\citenamefont {Tam}\ \emph {et~al.}(2022)\citenamefont {Tam}, \citenamefont {Choi}, \citenamefont {Ding}, \citenamefont {Agrestini}, \citenamefont {Nag}, \citenamefont {Wu}, \citenamefont {Huang}, \citenamefont {Luo}, \citenamefont {Gao}, \citenamefont {GarcÃ­a-FernÃ¡ndez}, \citenamefont {Qiao},\ and\ \citenamefont {Zhou}}]{Tam_Natmat2022}%
  \BibitemOpen
  \bibfield  {author} {\bibinfo {author} {\bibfnamefont {C.~C.}\ \bibnamefont {Tam}}, \bibinfo {author} {\bibfnamefont {J.}~\bibnamefont {Choi}}, \bibinfo {author} {\bibfnamefont {X.}~\bibnamefont {Ding}}, \bibinfo {author} {\bibfnamefont {S.}~\bibnamefont {Agrestini}}, \bibinfo {author} {\bibfnamefont {A.}~\bibnamefont {Nag}}, \bibinfo {author} {\bibfnamefont {M.}~\bibnamefont {Wu}}, \bibinfo {author} {\bibfnamefont {B.}~\bibnamefont {Huang}}, \bibinfo {author} {\bibfnamefont {H.}~\bibnamefont {Luo}}, \bibinfo {author} {\bibfnamefont {P.}~\bibnamefont {Gao}}, \bibinfo {author} {\bibfnamefont {M.}~\bibnamefont {GarcÃ­a-FernÃ¡ndez}}, \bibinfo {author} {\bibfnamefont {L.}~\bibnamefont {Qiao}},\ and\ \bibinfo {author} {\bibfnamefont {K.-J.}\ \bibnamefont {Zhou}},\ }\bibfield  {title} {\bibinfo {title} {Charge density waves in infinite-layer {NdNiO}$_2$ nickelates},\ }\href {https://doi.org/10.1038/s41563-022-01330-1} {\bibfield  {journal} {\bibinfo  {journal} {Nature Materials}\ }\textbf {\bibinfo {volume}
  {21}},\ \bibinfo {pages} {1116} (\bibinfo {year} {2022})}\BibitemShut {NoStop}%
\bibitem [{\citenamefont {Krieger}\ \emph {et~al.}(2022)\citenamefont {Krieger}, \citenamefont {Martinelli}, \citenamefont {Zeng}, \citenamefont {Chow}, \citenamefont {Kummer}, \citenamefont {Arpaia}, \citenamefont {Moretti~Sala}, \citenamefont {Brookes}, \citenamefont {Ariando}, \citenamefont {Viart}, \citenamefont {Salluzzo}, \citenamefont {Ghiringhelli},\ and\ \citenamefont {Preziosi}}]{Krieger_PRL2022}%
  \BibitemOpen
  \bibfield  {author} {\bibinfo {author} {\bibfnamefont {G.}~\bibnamefont {Krieger}}, \bibinfo {author} {\bibfnamefont {L.}~\bibnamefont {Martinelli}}, \bibinfo {author} {\bibfnamefont {S.}~\bibnamefont {Zeng}}, \bibinfo {author} {\bibfnamefont {L.~E.}\ \bibnamefont {Chow}}, \bibinfo {author} {\bibfnamefont {K.}~\bibnamefont {Kummer}}, \bibinfo {author} {\bibfnamefont {R.}~\bibnamefont {Arpaia}}, \bibinfo {author} {\bibfnamefont {M.}~\bibnamefont {Moretti~Sala}}, \bibinfo {author} {\bibfnamefont {N.~B.}\ \bibnamefont {Brookes}}, \bibinfo {author} {\bibfnamefont {A.}~\bibnamefont {Ariando}}, \bibinfo {author} {\bibfnamefont {N.}~\bibnamefont {Viart}}, \bibinfo {author} {\bibfnamefont {M.}~\bibnamefont {Salluzzo}}, \bibinfo {author} {\bibfnamefont {G.}~\bibnamefont {Ghiringhelli}},\ and\ \bibinfo {author} {\bibfnamefont {D.}~\bibnamefont {Preziosi}},\ }\bibfield  {title} {\bibinfo {title} {Charge and spin order dichotomy in {NdNiO}$_{2}$ driven by the capping layer},\ }\href
  {https://doi.org/10.1103/PhysRevLett.129.027002} {\bibfield  {journal} {\bibinfo  {journal} {Physical Review Letters}\ }\textbf {\bibinfo {volume} {129}},\ \bibinfo {pages} {027002} (\bibinfo {year} {2022})}\BibitemShut {NoStop}%
\bibitem [{\citenamefont {Yi}\ \emph {et~al.}(2018)\citenamefont {Yi}, \citenamefont {Frano}, \citenamefont {Lu}, \citenamefont {He}, \citenamefont {Wang}, \citenamefont {Frandsen}, \citenamefont {Kemper}, \citenamefont {Yu}, \citenamefont {Si}, \citenamefont {Wang}, \citenamefont {He}, \citenamefont {Hardy}, \citenamefont {Schweiss}, \citenamefont {Adelmann}, \citenamefont {Wolf}, \citenamefont {Hashimoto}, \citenamefont {Mo}, \citenamefont {Hussain}, \citenamefont {Le~Tacon}, \citenamefont {Bohmer}, \citenamefont {Lee}, \citenamefont {Shen}, \citenamefont {Meingast},\ and\ \citenamefont {Birgeneau}}]{Yi_PRL2018}%
  \BibitemOpen
  \bibfield  {author} {\bibinfo {author} {\bibfnamefont {M.}~\bibnamefont {Yi}}, \bibinfo {author} {\bibfnamefont {A.}~\bibnamefont {Frano}}, \bibinfo {author} {\bibfnamefont {D.~H.}\ \bibnamefont {Lu}}, \bibinfo {author} {\bibfnamefont {Y.}~\bibnamefont {He}}, \bibinfo {author} {\bibfnamefont {M.}~\bibnamefont {Wang}}, \bibinfo {author} {\bibfnamefont {B.~A.}\ \bibnamefont {Frandsen}}, \bibinfo {author} {\bibfnamefont {A.~F.}\ \bibnamefont {Kemper}}, \bibinfo {author} {\bibfnamefont {R.}~\bibnamefont {Yu}}, \bibinfo {author} {\bibfnamefont {Q.}~\bibnamefont {Si}}, \bibinfo {author} {\bibfnamefont {L.}~\bibnamefont {Wang}}, \bibinfo {author} {\bibfnamefont {M.}~\bibnamefont {He}}, \bibinfo {author} {\bibfnamefont {F.}~\bibnamefont {Hardy}}, \bibinfo {author} {\bibfnamefont {P.}~\bibnamefont {Schweiss}}, \bibinfo {author} {\bibfnamefont {P.}~\bibnamefont {Adelmann}}, \bibinfo {author} {\bibfnamefont {T.}~\bibnamefont {Wolf}}, \bibinfo {author} {\bibfnamefont {M.}~\bibnamefont {Hashimoto}}, \bibinfo {author}
  {\bibfnamefont {S.~K.}\ \bibnamefont {Mo}}, \bibinfo {author} {\bibfnamefont {Z.}~\bibnamefont {Hussain}}, \bibinfo {author} {\bibfnamefont {M.}~\bibnamefont {Le~Tacon}}, \bibinfo {author} {\bibfnamefont {A.~E.}\ \bibnamefont {Bohmer}}, \bibinfo {author} {\bibfnamefont {D.~H.}\ \bibnamefont {Lee}}, \bibinfo {author} {\bibfnamefont {Z.~X.}\ \bibnamefont {Shen}}, \bibinfo {author} {\bibfnamefont {C.}~\bibnamefont {Meingast}},\ and\ \bibinfo {author} {\bibfnamefont {R.~J.}\ \bibnamefont {Birgeneau}},\ }\bibfield  {title} {\bibinfo {title} {{Spectral Evidence for Emergent Order in {Ba}$_{1-x}${Na}$_x${Fe}$_2${As}$_2$}},\ }\href {https://doi.org/10.1103/PhysRevLett.121.127001} {\bibfield  {journal} {\bibinfo  {journal} {Physical Review Letters}\ }\textbf {\bibinfo {volume} {121}},\ \bibinfo {pages} {127001} (\bibinfo {year} {2018})}\BibitemShut {NoStop}%
\bibitem [{\citenamefont {Souliou}\ \emph {et~al.}(2022)\citenamefont {Souliou}, \citenamefont {Lacmann}, \citenamefont {Heid}, \citenamefont {Meingast}, \citenamefont {Frachet}, \citenamefont {Paolasini}, \citenamefont {Haghighirad}, \citenamefont {Merz}, \citenamefont {Bosak},\ and\ \citenamefont {Le~Tacon}}]{Souliou_PRL2022}%
  \BibitemOpen
  \bibfield  {author} {\bibinfo {author} {\bibfnamefont {S.~M.}\ \bibnamefont {Souliou}}, \bibinfo {author} {\bibfnamefont {T.}~\bibnamefont {Lacmann}}, \bibinfo {author} {\bibfnamefont {R.}~\bibnamefont {Heid}}, \bibinfo {author} {\bibfnamefont {C.}~\bibnamefont {Meingast}}, \bibinfo {author} {\bibfnamefont {M.}~\bibnamefont {Frachet}}, \bibinfo {author} {\bibfnamefont {L.}~\bibnamefont {Paolasini}}, \bibinfo {author} {\bibfnamefont {A.~A.}\ \bibnamefont {Haghighirad}}, \bibinfo {author} {\bibfnamefont {M.}~\bibnamefont {Merz}}, \bibinfo {author} {\bibfnamefont {A.}~\bibnamefont {Bosak}},\ and\ \bibinfo {author} {\bibfnamefont {M.}~\bibnamefont {Le~Tacon}},\ }\bibfield  {title} {\bibinfo {title} {Soft-phonon and charge-density-wave formation in nematic {BaNi}$_{2}${As}$_{2}$},\ }\href {https://doi.org/10.1103/PhysRevLett.129.247602} {\bibfield  {journal} {\bibinfo  {journal} {Physical Review Letters}\ }\textbf {\bibinfo {volume} {129}},\ \bibinfo {pages} {247602} (\bibinfo {year} {2022})}\BibitemShut
  {NoStop}%
\bibitem [{\citenamefont {Lacmann}\ \emph {et~al.}(2023)\citenamefont {Lacmann}, \citenamefont {Haghighirad}, \citenamefont {Souliou}, \citenamefont {Merz}, \citenamefont {Garbarino}, \citenamefont {Glazyrin}, \citenamefont {Heid},\ and\ \citenamefont {Le~Tacon}}]{Lacmann_PRB2023}%
  \BibitemOpen
  \bibfield  {author} {\bibinfo {author} {\bibfnamefont {T.}~\bibnamefont {Lacmann}}, \bibinfo {author} {\bibfnamefont {A.-A.}\ \bibnamefont {Haghighirad}}, \bibinfo {author} {\bibfnamefont {S.-M.}\ \bibnamefont {Souliou}}, \bibinfo {author} {\bibfnamefont {M.}~\bibnamefont {Merz}}, \bibinfo {author} {\bibfnamefont {G.}~\bibnamefont {Garbarino}}, \bibinfo {author} {\bibfnamefont {K.}~\bibnamefont {Glazyrin}}, \bibinfo {author} {\bibfnamefont {R.}~\bibnamefont {Heid}},\ and\ \bibinfo {author} {\bibfnamefont {M.}~\bibnamefont {Le~Tacon}},\ }\bibfield  {title} {\bibinfo {title} {High-pressure phase diagram of ${\mathrm{bani}}_{2}{\mathrm{as}}_{2}$: Unconventional charge density waves and structural phase transitions},\ }\href {https://doi.org/10.1103/PhysRevB.108.224115} {\bibfield  {journal} {\bibinfo  {journal} {Physical Review B}\ }\textbf {\bibinfo {volume} {108}},\ \bibinfo {pages} {224115} (\bibinfo {year} {2023})}\BibitemShut {NoStop}%
\bibitem [{\citenamefont {Leroux}\ \emph {et~al.}(2015)\citenamefont {Leroux}, \citenamefont {Errea}, \citenamefont {Le~Tacon}, \citenamefont {Souliou}, \citenamefont {Garbarino}, \citenamefont {Cario}, \citenamefont {Bosak}, \citenamefont {Mauri}, \citenamefont {Calandra},\ and\ \citenamefont {Rodiere}}]{Leroux_PRB2015}%
  \BibitemOpen
  \bibfield  {author} {\bibinfo {author} {\bibfnamefont {M.}~\bibnamefont {Leroux}}, \bibinfo {author} {\bibfnamefont {I.}~\bibnamefont {Errea}}, \bibinfo {author} {\bibfnamefont {M.}~\bibnamefont {Le~Tacon}}, \bibinfo {author} {\bibfnamefont {S.~M.}\ \bibnamefont {Souliou}}, \bibinfo {author} {\bibfnamefont {G.}~\bibnamefont {Garbarino}}, \bibinfo {author} {\bibfnamefont {L.}~\bibnamefont {Cario}}, \bibinfo {author} {\bibfnamefont {A.}~\bibnamefont {Bosak}}, \bibinfo {author} {\bibfnamefont {F.}~\bibnamefont {Mauri}}, \bibinfo {author} {\bibfnamefont {M.}~\bibnamefont {Calandra}},\ and\ \bibinfo {author} {\bibfnamefont {P.}~\bibnamefont {Rodiere}},\ }\bibfield  {title} {\bibinfo {title} {Strong anharmonicity induces quantum melting of charge density wave in 2h-nbse2 under pressure},\ }\href {https://doi.org/10.1103/PhysRevB.92.140303} {\bibfield  {journal} {\bibinfo  {journal} {Physical Review B}\ }\textbf {\bibinfo {volume} {92}},\ \bibinfo {pages} {140303} (\bibinfo {year} {2015})}\BibitemShut {NoStop}%
\bibitem [{\citenamefont {Aishwarya}\ \emph {et~al.}(2023)\citenamefont {Aishwarya}, \citenamefont {May-Mann}, \citenamefont {Raghavan}, \citenamefont {Nie}, \citenamefont {Romanelli}, \citenamefont {Ran}, \citenamefont {Saha}, \citenamefont {Paglione}, \citenamefont {Butch}, \citenamefont {Fradkin},\ and\ \citenamefont {Madhavan}}]{Aishwarya_Nature2023}%
  \BibitemOpen
  \bibfield  {author} {\bibinfo {author} {\bibfnamefont {A.}~\bibnamefont {Aishwarya}}, \bibinfo {author} {\bibfnamefont {J.}~\bibnamefont {May-Mann}}, \bibinfo {author} {\bibfnamefont {A.}~\bibnamefont {Raghavan}}, \bibinfo {author} {\bibfnamefont {L.}~\bibnamefont {Nie}}, \bibinfo {author} {\bibfnamefont {M.}~\bibnamefont {Romanelli}}, \bibinfo {author} {\bibfnamefont {S.}~\bibnamefont {Ran}}, \bibinfo {author} {\bibfnamefont {S.~R.}\ \bibnamefont {Saha}}, \bibinfo {author} {\bibfnamefont {J.}~\bibnamefont {Paglione}}, \bibinfo {author} {\bibfnamefont {N.~P.}\ \bibnamefont {Butch}}, \bibinfo {author} {\bibfnamefont {E.}~\bibnamefont {Fradkin}},\ and\ \bibinfo {author} {\bibfnamefont {V.}~\bibnamefont {Madhavan}},\ }\bibfield  {title} {\bibinfo {title} {Magnetic-field-sensitive charge density waves in the superconductor ute2},\ }\href {https://doi.org/10.1038/s41586-023-06005-8} {\bibfield  {journal} {\bibinfo  {journal} {Nature}\ }\textbf {\bibinfo {volume} {618}},\ \bibinfo {pages} {928} (\bibinfo {year}
  {2023})}\BibitemShut {NoStop}%
\bibitem [{\citenamefont {Ortiz}\ \emph {et~al.}(2020)\citenamefont {Ortiz}, \citenamefont {Teicher}, \citenamefont {Hu}, \citenamefont {Zuo}, \citenamefont {Sarte}, \citenamefont {Schueller}, \citenamefont {Abeykoon}, \citenamefont {Krogstad}, \citenamefont {Rosenkranz}, \citenamefont {Osborn}, \citenamefont {Seshadri}, \citenamefont {Balents}, \citenamefont {He},\ and\ \citenamefont {Wilson}}]{Ortiz_PRL2020}%
  \BibitemOpen
  \bibfield  {author} {\bibinfo {author} {\bibfnamefont {B.~R.}\ \bibnamefont {Ortiz}}, \bibinfo {author} {\bibfnamefont {S.~M.~L.}\ \bibnamefont {Teicher}}, \bibinfo {author} {\bibfnamefont {Y.}~\bibnamefont {Hu}}, \bibinfo {author} {\bibfnamefont {J.~L.}\ \bibnamefont {Zuo}}, \bibinfo {author} {\bibfnamefont {P.~M.}\ \bibnamefont {Sarte}}, \bibinfo {author} {\bibfnamefont {E.~C.}\ \bibnamefont {Schueller}}, \bibinfo {author} {\bibfnamefont {A.~M.~M.}\ \bibnamefont {Abeykoon}}, \bibinfo {author} {\bibfnamefont {M.~J.}\ \bibnamefont {Krogstad}}, \bibinfo {author} {\bibfnamefont {S.}~\bibnamefont {Rosenkranz}}, \bibinfo {author} {\bibfnamefont {R.}~\bibnamefont {Osborn}}, \bibinfo {author} {\bibfnamefont {R.}~\bibnamefont {Seshadri}}, \bibinfo {author} {\bibfnamefont {L.}~\bibnamefont {Balents}}, \bibinfo {author} {\bibfnamefont {J.}~\bibnamefont {He}},\ and\ \bibinfo {author} {\bibfnamefont {S.~D.}\ \bibnamefont {Wilson}},\ }\bibfield  {title} {\bibinfo {title} {{CsV}$_{3}${Sb}$_{5}$: A ${\mathbb{z}}_{2}$
  topological kagome metal with a superconducting ground state},\ }\href {https://doi.org/10.1103/PhysRevLett.125.247002} {\bibfield  {journal} {\bibinfo  {journal} {Physical Review Letters}\ }\textbf {\bibinfo {volume} {125}},\ \bibinfo {pages} {247002} (\bibinfo {year} {2020})}\BibitemShut {NoStop}%
\bibitem [{\citenamefont {Li}\ \emph {et~al.}(2021)\citenamefont {Li}, \citenamefont {Zhang}, \citenamefont {Yilmaz}, \citenamefont {Pai}, \citenamefont {Marvinney}, \citenamefont {Said}, \citenamefont {Yin}, \citenamefont {Gong}, \citenamefont {Tu}, \citenamefont {Vescovo}, \citenamefont {Nelson}, \citenamefont {Moore}, \citenamefont {Murakami}, \citenamefont {Lei}, \citenamefont {Lee}, \citenamefont {Lawrie},\ and\ \citenamefont {Miao}}]{Li_PRX2021}%
  \BibitemOpen
  \bibfield  {author} {\bibinfo {author} {\bibfnamefont {H.}~\bibnamefont {Li}}, \bibinfo {author} {\bibfnamefont {T.~T.}\ \bibnamefont {Zhang}}, \bibinfo {author} {\bibfnamefont {T.}~\bibnamefont {Yilmaz}}, \bibinfo {author} {\bibfnamefont {Y.~Y.}\ \bibnamefont {Pai}}, \bibinfo {author} {\bibfnamefont {C.~E.}\ \bibnamefont {Marvinney}}, \bibinfo {author} {\bibfnamefont {A.}~\bibnamefont {Said}}, \bibinfo {author} {\bibfnamefont {Q.~W.}\ \bibnamefont {Yin}}, \bibinfo {author} {\bibfnamefont {C.~S.}\ \bibnamefont {Gong}}, \bibinfo {author} {\bibfnamefont {Z.~J.}\ \bibnamefont {Tu}}, \bibinfo {author} {\bibfnamefont {E.}~\bibnamefont {Vescovo}}, \bibinfo {author} {\bibfnamefont {C.~S.}\ \bibnamefont {Nelson}}, \bibinfo {author} {\bibfnamefont {R.~G.}\ \bibnamefont {Moore}}, \bibinfo {author} {\bibfnamefont {S.}~\bibnamefont {Murakami}}, \bibinfo {author} {\bibfnamefont {H.~C.}\ \bibnamefont {Lei}}, \bibinfo {author} {\bibfnamefont {H.~N.}\ \bibnamefont {Lee}}, \bibinfo {author} {\bibfnamefont {B.~J.}\
  \bibnamefont {Lawrie}},\ and\ \bibinfo {author} {\bibfnamefont {H.}~\bibnamefont {Miao}},\ }\bibfield  {title} {\bibinfo {title} {Observation of unconventional charge density wave without acoustic phonon anomaly in kagome superconductors {AV}$_{3}${Sb}$_{5}$ ({A}={Rb}, {Cs})},\ }\href {https://doi.org/10.1103/PhysRevX.11.031050} {\bibfield  {journal} {\bibinfo  {journal} {Physical Review X}\ }\textbf {\bibinfo {volume} {11}},\ \bibinfo {pages} {031050} (\bibinfo {year} {2021})}\BibitemShut {NoStop}%
\bibitem [{\citenamefont {Subires}\ \emph {et~al.}(2023)\citenamefont {Subires}, \citenamefont {Korshunov}, \citenamefont {Said}, \citenamefont {SÃ¡nchez}, \citenamefont {Ortiz}, \citenamefont {Wilson}, \citenamefont {Bosak},\ and\ \citenamefont {Blanco-Canosa}}]{Subires_NatCom2023}%
  \BibitemOpen
  \bibfield  {author} {\bibinfo {author} {\bibfnamefont {D.}~\bibnamefont {Subires}}, \bibinfo {author} {\bibfnamefont {A.}~\bibnamefont {Korshunov}}, \bibinfo {author} {\bibfnamefont {A.~H.}\ \bibnamefont {Said}}, \bibinfo {author} {\bibfnamefont {L.}~\bibnamefont {SÃ¡nchez}}, \bibinfo {author} {\bibfnamefont {B.~R.}\ \bibnamefont {Ortiz}}, \bibinfo {author} {\bibfnamefont {S.~D.}\ \bibnamefont {Wilson}}, \bibinfo {author} {\bibfnamefont {A.}~\bibnamefont {Bosak}},\ and\ \bibinfo {author} {\bibfnamefont {S.}~\bibnamefont {Blanco-Canosa}},\ }\bibfield  {title} {\bibinfo {title} {Order-disorder charge density wave instability in the kagome metal {(Cs,Rb)V}$_{3}${Sb}$_{5}$},\ }\href {https://doi.org/10.1038/s41467-023-36668-w} {\bibfield  {journal} {\bibinfo  {journal} {Nature Communications}\ }\textbf {\bibinfo {volume} {14}},\ \bibinfo {pages} {1015} (\bibinfo {year} {2023})}\BibitemShut {NoStop}%
\bibitem [{\citenamefont {Guo}\ \emph {et~al.}(2022)\citenamefont {Guo}, \citenamefont {Putzke}, \citenamefont {Konyzheva}, \citenamefont {Huang}, \citenamefont {Gutierrez-Amigo}, \citenamefont {Errea}, \citenamefont {Chen}, \citenamefont {Vergniory}, \citenamefont {Felser}, \citenamefont {Fischer}, \citenamefont {Neupert},\ and\ \citenamefont {Moll}}]{Guo_Nature2022}%
  \BibitemOpen
  \bibfield  {author} {\bibinfo {author} {\bibfnamefont {C.}~\bibnamefont {Guo}}, \bibinfo {author} {\bibfnamefont {C.}~\bibnamefont {Putzke}}, \bibinfo {author} {\bibfnamefont {S.}~\bibnamefont {Konyzheva}}, \bibinfo {author} {\bibfnamefont {X.}~\bibnamefont {Huang}}, \bibinfo {author} {\bibfnamefont {M.}~\bibnamefont {Gutierrez-Amigo}}, \bibinfo {author} {\bibfnamefont {I.}~\bibnamefont {Errea}}, \bibinfo {author} {\bibfnamefont {D.}~\bibnamefont {Chen}}, \bibinfo {author} {\bibfnamefont {M.~G.}\ \bibnamefont {Vergniory}}, \bibinfo {author} {\bibfnamefont {C.}~\bibnamefont {Felser}}, \bibinfo {author} {\bibfnamefont {M.~H.}\ \bibnamefont {Fischer}}, \bibinfo {author} {\bibfnamefont {T.}~\bibnamefont {Neupert}},\ and\ \bibinfo {author} {\bibfnamefont {P.~J.~W.}\ \bibnamefont {Moll}},\ }\bibfield  {title} {\bibinfo {title} {Switchable chiral transport in charge-ordered kagome metal csv3sb5},\ }\href {https://doi.org/10.1038/s41586-022-05127-9} {\bibfield  {journal} {\bibinfo  {journal} {Nature}\ }\textbf
  {\bibinfo {volume} {611}},\ \bibinfo {pages} {461} (\bibinfo {year} {2022})}\BibitemShut {NoStop}%
\bibitem [{\citenamefont {Nie}\ \emph {et~al.}(2022)\citenamefont {Nie}, \citenamefont {Sun}, \citenamefont {Ma}, \citenamefont {Song}, \citenamefont {Zheng}, \citenamefont {Liang}, \citenamefont {Wu}, \citenamefont {Yu}, \citenamefont {Li}, \citenamefont {Shan}, \citenamefont {Zhao}, \citenamefont {Li}, \citenamefont {Kang}, \citenamefont {Wu}, \citenamefont {Zhou}, \citenamefont {Liu}, \citenamefont {Xiang}, \citenamefont {Ying}, \citenamefont {Wang}, \citenamefont {Wu},\ and\ \citenamefont {Chen}}]{Nie_Nature2022}%
  \BibitemOpen
  \bibfield  {author} {\bibinfo {author} {\bibfnamefont {L.}~\bibnamefont {Nie}}, \bibinfo {author} {\bibfnamefont {K.}~\bibnamefont {Sun}}, \bibinfo {author} {\bibfnamefont {W.}~\bibnamefont {Ma}}, \bibinfo {author} {\bibfnamefont {D.}~\bibnamefont {Song}}, \bibinfo {author} {\bibfnamefont {L.}~\bibnamefont {Zheng}}, \bibinfo {author} {\bibfnamefont {Z.}~\bibnamefont {Liang}}, \bibinfo {author} {\bibfnamefont {P.}~\bibnamefont {Wu}}, \bibinfo {author} {\bibfnamefont {F.}~\bibnamefont {Yu}}, \bibinfo {author} {\bibfnamefont {J.}~\bibnamefont {Li}}, \bibinfo {author} {\bibfnamefont {M.}~\bibnamefont {Shan}}, \bibinfo {author} {\bibfnamefont {D.}~\bibnamefont {Zhao}}, \bibinfo {author} {\bibfnamefont {S.}~\bibnamefont {Li}}, \bibinfo {author} {\bibfnamefont {B.}~\bibnamefont {Kang}}, \bibinfo {author} {\bibfnamefont {Z.}~\bibnamefont {Wu}}, \bibinfo {author} {\bibfnamefont {Y.}~\bibnamefont {Zhou}}, \bibinfo {author} {\bibfnamefont {K.}~\bibnamefont {Liu}}, \bibinfo {author} {\bibfnamefont {Z.}~\bibnamefont
  {Xiang}}, \bibinfo {author} {\bibfnamefont {J.}~\bibnamefont {Ying}}, \bibinfo {author} {\bibfnamefont {Z.}~\bibnamefont {Wang}}, \bibinfo {author} {\bibfnamefont {T.}~\bibnamefont {Wu}},\ and\ \bibinfo {author} {\bibfnamefont {X.}~\bibnamefont {Chen}},\ }\bibfield  {title} {\bibinfo {title} {Charge-density-wave-driven electronic nematicity in a kagome superconductor},\ }\href {https://doi.org/10.1038/s41586-022-04493-8} {\bibfield  {journal} {\bibinfo  {journal} {Nature}\ }\textbf {\bibinfo {volume} {604}},\ \bibinfo {pages} {59} (\bibinfo {year} {2022})}\BibitemShut {NoStop}%
\bibitem [{\citenamefont {Zheng}\ \emph {et~al.}(2022)\citenamefont {Zheng}, \citenamefont {Wu}, \citenamefont {Yang}, \citenamefont {Nie}, \citenamefont {Shan}, \citenamefont {Sun}, \citenamefont {Song}, \citenamefont {Yu}, \citenamefont {Li}, \citenamefont {Zhao}, \citenamefont {Li}, \citenamefont {Kang}, \citenamefont {Zhou}, \citenamefont {Liu}, \citenamefont {Xiang}, \citenamefont {Ying}, \citenamefont {Wang}, \citenamefont {Wu},\ and\ \citenamefont {Chen}}]{Zheng_Nature2022}%
  \BibitemOpen
  \bibfield  {author} {\bibinfo {author} {\bibfnamefont {L.}~\bibnamefont {Zheng}}, \bibinfo {author} {\bibfnamefont {Z.}~\bibnamefont {Wu}}, \bibinfo {author} {\bibfnamefont {Y.}~\bibnamefont {Yang}}, \bibinfo {author} {\bibfnamefont {L.}~\bibnamefont {Nie}}, \bibinfo {author} {\bibfnamefont {M.}~\bibnamefont {Shan}}, \bibinfo {author} {\bibfnamefont {K.}~\bibnamefont {Sun}}, \bibinfo {author} {\bibfnamefont {D.}~\bibnamefont {Song}}, \bibinfo {author} {\bibfnamefont {F.}~\bibnamefont {Yu}}, \bibinfo {author} {\bibfnamefont {J.}~\bibnamefont {Li}}, \bibinfo {author} {\bibfnamefont {D.}~\bibnamefont {Zhao}}, \bibinfo {author} {\bibfnamefont {S.}~\bibnamefont {Li}}, \bibinfo {author} {\bibfnamefont {B.}~\bibnamefont {Kang}}, \bibinfo {author} {\bibfnamefont {Y.}~\bibnamefont {Zhou}}, \bibinfo {author} {\bibfnamefont {K.}~\bibnamefont {Liu}}, \bibinfo {author} {\bibfnamefont {Z.}~\bibnamefont {Xiang}}, \bibinfo {author} {\bibfnamefont {J.}~\bibnamefont {Ying}}, \bibinfo {author} {\bibfnamefont {Z.}~\bibnamefont
  {Wang}}, \bibinfo {author} {\bibfnamefont {T.}~\bibnamefont {Wu}},\ and\ \bibinfo {author} {\bibfnamefont {X.}~\bibnamefont {Chen}},\ }\bibfield  {title} {\bibinfo {title} {Emergent charge order in pressurized kagome superconductor {CsV}$_{3}${Sb}$_{5}$},\ }\href {https://doi.org/10.1038/s41586-022-05351-3} {\bibfield  {journal} {\bibinfo  {journal} {Nature}\ }\textbf {\bibinfo {volume} {611}},\ \bibinfo {pages} {682} (\bibinfo {year} {2022})}\BibitemShut {NoStop}%
\bibitem [{\citenamefont {Asaba}\ \emph {et~al.}(2024)\citenamefont {Asaba}, \citenamefont {Onishi}, \citenamefont {Kageyama}, \citenamefont {Kiyosue}, \citenamefont {Ohtsuka}, \citenamefont {Suetsugu}, \citenamefont {Kohsaka}, \citenamefont {Gaggl}, \citenamefont {Kasahara}, \citenamefont {Murayama}, \citenamefont {Hashimoto}, \citenamefont {Tazai}, \citenamefont {Kontani}, \citenamefont {Ortiz}, \citenamefont {Wilson}, \citenamefont {Li}, \citenamefont {Wen}, \citenamefont {Shibauchi},\ and\ \citenamefont {Matsuda}}]{Asaba_2023}%
  \BibitemOpen
  \bibfield  {author} {\bibinfo {author} {\bibfnamefont {T.}~\bibnamefont {Asaba}}, \bibinfo {author} {\bibfnamefont {A.}~\bibnamefont {Onishi}}, \bibinfo {author} {\bibfnamefont {Y.}~\bibnamefont {Kageyama}}, \bibinfo {author} {\bibfnamefont {T.}~\bibnamefont {Kiyosue}}, \bibinfo {author} {\bibfnamefont {K.}~\bibnamefont {Ohtsuka}}, \bibinfo {author} {\bibfnamefont {S.}~\bibnamefont {Suetsugu}}, \bibinfo {author} {\bibfnamefont {Y.}~\bibnamefont {Kohsaka}}, \bibinfo {author} {\bibfnamefont {T.}~\bibnamefont {Gaggl}}, \bibinfo {author} {\bibfnamefont {Y.}~\bibnamefont {Kasahara}}, \bibinfo {author} {\bibfnamefont {H.}~\bibnamefont {Murayama}}, \bibinfo {author} {\bibfnamefont {K.}~\bibnamefont {Hashimoto}}, \bibinfo {author} {\bibfnamefont {R.}~\bibnamefont {Tazai}}, \bibinfo {author} {\bibfnamefont {H.}~\bibnamefont {Kontani}}, \bibinfo {author} {\bibfnamefont {B.~R.}\ \bibnamefont {Ortiz}}, \bibinfo {author} {\bibfnamefont {S.~D.}\ \bibnamefont {Wilson}}, \bibinfo {author} {\bibfnamefont {Q.}~\bibnamefont
  {Li}}, \bibinfo {author} {\bibfnamefont {H.~H.}\ \bibnamefont {Wen}}, \bibinfo {author} {\bibfnamefont {T.}~\bibnamefont {Shibauchi}},\ and\ \bibinfo {author} {\bibfnamefont {Y.}~\bibnamefont {Matsuda}},\ }\bibfield  {title} {\bibinfo {title} {{Evidence for an odd-parity nematic phase above the charge-density-wave transition in a kagome metal}},\ }\href {https://doi.org/10.1038/s41567-023-02272-4} {\bibfield  {journal} {\bibinfo  {journal} {Nature Physics}\ }\textbf {\bibinfo {volume} {20}},\ \bibinfo {pages} {40} (\bibinfo {year} {2024})}\BibitemShut {NoStop}%
\bibitem [{\citenamefont {Frachet}\ \emph {et~al.}(2024)\citenamefont {Frachet}, \citenamefont {Wang}, \citenamefont {Xia}, \citenamefont {Guo}, \citenamefont {He}, \citenamefont {Maraytta}, \citenamefont {Heid}, \citenamefont {Haghighirad}, \citenamefont {Merz}, \citenamefont {Meingast},\ and\ \citenamefont {Hardy}}]{Frachet_2023}%
  \BibitemOpen
  \bibfield  {author} {\bibinfo {author} {\bibfnamefont {M.}~\bibnamefont {Frachet}}, \bibinfo {author} {\bibfnamefont {L.}~\bibnamefont {Wang}}, \bibinfo {author} {\bibfnamefont {W.}~\bibnamefont {Xia}}, \bibinfo {author} {\bibfnamefont {Y.}~\bibnamefont {Guo}}, \bibinfo {author} {\bibfnamefont {M.}~\bibnamefont {He}}, \bibinfo {author} {\bibfnamefont {N.}~\bibnamefont {Maraytta}}, \bibinfo {author} {\bibfnamefont {R.}~\bibnamefont {Heid}}, \bibinfo {author} {\bibfnamefont {A.-A.}\ \bibnamefont {Haghighirad}}, \bibinfo {author} {\bibfnamefont {M.}~\bibnamefont {Merz}}, \bibinfo {author} {\bibfnamefont {C.}~\bibnamefont {Meingast}},\ and\ \bibinfo {author} {\bibfnamefont {F.}~\bibnamefont {Hardy}},\ }\bibfield  {title} {\bibinfo {title} {Colossal $c$-axis response and lack of rotational symmetry breaking within the kagome planes of the {CsV}$_3${Sb}$_5$ superconductor},\ }\href {https://doi.org/10.1103/PhysRevLett.132.186001} {\bibfield  {journal} {\bibinfo  {journal} {Phys. Rev. Lett.}\ }\textbf {\bibinfo
  {volume} {132}},\ \bibinfo {pages} {186001} (\bibinfo {year} {2024})}\BibitemShut {NoStop}%
\bibitem [{\citenamefont {Liu}\ \emph {et~al.}(2024)\citenamefont {Liu}, \citenamefont {Shi}, \citenamefont {Jiang}, \citenamefont {Rosenberg}, \citenamefont {DeStefano}, \citenamefont {Liu}, \citenamefont {Hu}, \citenamefont {Zhao}, \citenamefont {Wang}, \citenamefont {Yao}, \citenamefont {Graf}, \citenamefont {Dai}, \citenamefont {Yang}, \citenamefont {Xu},\ and\ \citenamefont {Chu}}]{Liu_2023}%
  \BibitemOpen
  \bibfield  {author} {\bibinfo {author} {\bibfnamefont {Z.}~\bibnamefont {Liu}}, \bibinfo {author} {\bibfnamefont {Y.}~\bibnamefont {Shi}}, \bibinfo {author} {\bibfnamefont {Q.}~\bibnamefont {Jiang}}, \bibinfo {author} {\bibfnamefont {E.~W.}\ \bibnamefont {Rosenberg}}, \bibinfo {author} {\bibfnamefont {J.~M.}\ \bibnamefont {DeStefano}}, \bibinfo {author} {\bibfnamefont {J.}~\bibnamefont {Liu}}, \bibinfo {author} {\bibfnamefont {C.}~\bibnamefont {Hu}}, \bibinfo {author} {\bibfnamefont {Y.}~\bibnamefont {Zhao}}, \bibinfo {author} {\bibfnamefont {Z.}~\bibnamefont {Wang}}, \bibinfo {author} {\bibfnamefont {Y.}~\bibnamefont {Yao}}, \bibinfo {author} {\bibfnamefont {D.}~\bibnamefont {Graf}}, \bibinfo {author} {\bibfnamefont {P.}~\bibnamefont {Dai}}, \bibinfo {author} {\bibfnamefont {J.}~\bibnamefont {Yang}}, \bibinfo {author} {\bibfnamefont {X.}~\bibnamefont {Xu}},\ and\ \bibinfo {author} {\bibfnamefont {J.-H.}\ \bibnamefont {Chu}},\ }\bibfield  {title} {\bibinfo {title} {{Absence of ${E}_{2g}$ Nematic Instability
  and Dominant ${A}_{1g}$ Response in the Kagome Metal ${\mathrm{CsV}}_{3}{\mathrm{Sb}}_{5}$}},\ }\href {https://doi.org/10.1103/PhysRevX.14.031015} {\bibfield  {journal} {\bibinfo  {journal} {Phys. Rev. X}\ }\textbf {\bibinfo {volume} {14}},\ \bibinfo {pages} {031015} (\bibinfo {year} {2024})}\BibitemShut {NoStop}%
\bibitem [{\citenamefont {Zhong}\ \emph {et~al.}(2023)\citenamefont {Zhong}, \citenamefont {Li}, \citenamefont {Liu}, \citenamefont {Dong}, \citenamefont {Aido}, \citenamefont {Arai}, \citenamefont {Li}, \citenamefont {Zhang}, \citenamefont {Shi}, \citenamefont {Wang}, \citenamefont {Shin}, \citenamefont {Lee}, \citenamefont {Miao}, \citenamefont {Kondo},\ and\ \citenamefont {Okazaki}}]{Zhong_NatCom2023}%
  \BibitemOpen
  \bibfield  {author} {\bibinfo {author} {\bibfnamefont {Y.}~\bibnamefont {Zhong}}, \bibinfo {author} {\bibfnamefont {S.}~\bibnamefont {Li}}, \bibinfo {author} {\bibfnamefont {H.}~\bibnamefont {Liu}}, \bibinfo {author} {\bibfnamefont {Y.}~\bibnamefont {Dong}}, \bibinfo {author} {\bibfnamefont {K.}~\bibnamefont {Aido}}, \bibinfo {author} {\bibfnamefont {Y.}~\bibnamefont {Arai}}, \bibinfo {author} {\bibfnamefont {H.}~\bibnamefont {Li}}, \bibinfo {author} {\bibfnamefont {W.}~\bibnamefont {Zhang}}, \bibinfo {author} {\bibfnamefont {Y.}~\bibnamefont {Shi}}, \bibinfo {author} {\bibfnamefont {Z.}~\bibnamefont {Wang}}, \bibinfo {author} {\bibfnamefont {S.}~\bibnamefont {Shin}}, \bibinfo {author} {\bibfnamefont {H.~N.}\ \bibnamefont {Lee}}, \bibinfo {author} {\bibfnamefont {H.}~\bibnamefont {Miao}}, \bibinfo {author} {\bibfnamefont {T.}~\bibnamefont {Kondo}},\ and\ \bibinfo {author} {\bibfnamefont {K.}~\bibnamefont {Okazaki}},\ }\bibfield  {title} {\bibinfo {title} {Testing electron-phonon coupling for the
  superconductivity in kagome metal {CsV}$_{3}${Sb}$_{5}$},\ }\href {https://doi.org/10.1038/s41467-023-37605-7} {\bibfield  {journal} {\bibinfo  {journal} {Nature Communications}\ }\textbf {\bibinfo {volume} {14}},\ \bibinfo {pages} {1945} (\bibinfo {year} {2023})}\BibitemShut {NoStop}%
\bibitem [{\citenamefont {Ritz}\ \emph {et~al.}(2023)\citenamefont {Ritz}, \citenamefont {R\o{}ising}, \citenamefont {Christensen}, \citenamefont {Birol}, \citenamefont {Andersen},\ and\ \citenamefont {Fernandes}}]{Ritz_PRB2023}%
  \BibitemOpen
  \bibfield  {author} {\bibinfo {author} {\bibfnamefont {E.~T.}\ \bibnamefont {Ritz}}, \bibinfo {author} {\bibfnamefont {H.~S.}\ \bibnamefont {R\o{}ising}}, \bibinfo {author} {\bibfnamefont {M.~H.}\ \bibnamefont {Christensen}}, \bibinfo {author} {\bibfnamefont {T.}~\bibnamefont {Birol}}, \bibinfo {author} {\bibfnamefont {B.~M.}\ \bibnamefont {Andersen}},\ and\ \bibinfo {author} {\bibfnamefont {R.~M.}\ \bibnamefont {Fernandes}},\ }\bibfield  {title} {\bibinfo {title} {Superconductivity from orbital-selective electron-phonon coupling in {AV}$_3${Sb}$_5$},\ }\href {https://doi.org/10.1103/PhysRevB.108.L100510} {\bibfield  {journal} {\bibinfo  {journal} {Phys. Rev. B}\ }\textbf {\bibinfo {volume} {108}},\ \bibinfo {pages} {L100510} (\bibinfo {year} {2023})}\BibitemShut {NoStop}%
\bibitem [{\citenamefont {Souliou}\ \emph {et~al.}(2020)\citenamefont {Souliou}, \citenamefont {Bosak}, \citenamefont {Garbarino},\ and\ \citenamefont {Tacon}}]{Souliou_2020}%
  \BibitemOpen
  \bibfield  {author} {\bibinfo {author} {\bibfnamefont {S.~M.}\ \bibnamefont {Souliou}}, \bibinfo {author} {\bibfnamefont {A.}~\bibnamefont {Bosak}}, \bibinfo {author} {\bibfnamefont {G.}~\bibnamefont {Garbarino}},\ and\ \bibinfo {author} {\bibfnamefont {M.~L.}\ \bibnamefont {Tacon}},\ }\bibfield  {title} {\bibinfo {title} {Inelastic x-ray scattering studies of phonon dispersions in superconductors at high pressures},\ }\href {https://doi.org/10.1088/1361-6668/abbdc3} {\bibfield  {journal} {\bibinfo  {journal} {Superconductor Science and Technology}\ }\textbf {\bibinfo {volume} {33}},\ \bibinfo {pages} {124004} (\bibinfo {year} {2020})}\BibitemShut {NoStop}%
\bibitem [{\citenamefont {Pimenta~Martins}\ \emph {et~al.}(2023)\citenamefont {Pimenta~Martins}, \citenamefont {Comin}, \citenamefont {Matos}, \citenamefont {Mazzoni}, \citenamefont {Neves},\ and\ \citenamefont {Yankowitz}}]{Pimenta_APR2023}%
  \BibitemOpen
  \bibfield  {author} {\bibinfo {author} {\bibfnamefont {L.~G.}\ \bibnamefont {Pimenta~Martins}}, \bibinfo {author} {\bibfnamefont {R.}~\bibnamefont {Comin}}, \bibinfo {author} {\bibfnamefont {M.~J.~S.}\ \bibnamefont {Matos}}, \bibinfo {author} {\bibfnamefont {M.~S.~C.}\ \bibnamefont {Mazzoni}}, \bibinfo {author} {\bibfnamefont {B.~R.~A.}\ \bibnamefont {Neves}},\ and\ \bibinfo {author} {\bibfnamefont {M.}~\bibnamefont {Yankowitz}},\ }\bibfield  {title} {\bibinfo {title} {High-pressure studies of atomically thin van der waals materials},\ }\href {https://doi.org/10.1063/5.0123283} {\bibfield  {journal} {\bibinfo  {journal} {Applied Physics Reviews}\ }\textbf {\bibinfo {volume} {10}},\ \bibinfo {pages} {011313} (\bibinfo {year} {2023})}\BibitemShut {NoStop}%
\bibitem [{\citenamefont {Mao}\ \emph {et~al.}(2018)\citenamefont {Mao}, \citenamefont {Chen}, \citenamefont {Ding}, \citenamefont {Li},\ and\ \citenamefont {Wang}}]{Mao_RPM2018}%
  \BibitemOpen
  \bibfield  {author} {\bibinfo {author} {\bibfnamefont {H.-K.}\ \bibnamefont {Mao}}, \bibinfo {author} {\bibfnamefont {X.-J.}\ \bibnamefont {Chen}}, \bibinfo {author} {\bibfnamefont {Y.}~\bibnamefont {Ding}}, \bibinfo {author} {\bibfnamefont {B.}~\bibnamefont {Li}},\ and\ \bibinfo {author} {\bibfnamefont {L.}~\bibnamefont {Wang}},\ }\bibfield  {title} {\bibinfo {title} {Solids, liquids, and gases under high pressure},\ }\href {https://doi.org/10.1103/RevModPhys.90.015007} {\bibfield  {journal} {\bibinfo  {journal} {Rev. Mod. Phys.}\ }\textbf {\bibinfo {volume} {90}},\ \bibinfo {pages} {015007} (\bibinfo {year} {2018})}\BibitemShut {NoStop}%
\bibitem [{\citenamefont {Yu}\ \emph {et~al.}(2021)\citenamefont {Yu}, \citenamefont {Ma}, \citenamefont {Zhuo}, \citenamefont {Liu}, \citenamefont {Wen}, \citenamefont {Lei}, \citenamefont {Ying},\ and\ \citenamefont {Chen}}]{Yu_NatCom2021}%
  \BibitemOpen
  \bibfield  {author} {\bibinfo {author} {\bibfnamefont {F.~H.}\ \bibnamefont {Yu}}, \bibinfo {author} {\bibfnamefont {D.~H.}\ \bibnamefont {Ma}}, \bibinfo {author} {\bibfnamefont {W.~Z.}\ \bibnamefont {Zhuo}}, \bibinfo {author} {\bibfnamefont {S.~Q.}\ \bibnamefont {Liu}}, \bibinfo {author} {\bibfnamefont {X.~K.}\ \bibnamefont {Wen}}, \bibinfo {author} {\bibfnamefont {B.}~\bibnamefont {Lei}}, \bibinfo {author} {\bibfnamefont {J.~J.}\ \bibnamefont {Ying}},\ and\ \bibinfo {author} {\bibfnamefont {X.~H.}\ \bibnamefont {Chen}},\ }\bibfield  {title} {\bibinfo {title} {Unusual competition of superconductivity and charge-density-wave state in a compressed topological kagome metal},\ }\href {https://doi.org/10.1038/s41467-021-23928-w} {\bibfield  {journal} {\bibinfo  {journal} {Nature Communications}\ }\textbf {\bibinfo {volume} {12}},\ \bibinfo {pages} {3645} (\bibinfo {year} {2021})}\BibitemShut {NoStop}%
\bibitem [{\citenamefont {Feng}\ \emph {et~al.}(2023)\citenamefont {Feng}, \citenamefont {Zhao}, \citenamefont {Luo}, \citenamefont {Yang}, \citenamefont {Fang}, \citenamefont {Yang}, \citenamefont {Gao}, \citenamefont {Zhou},\ and\ \citenamefont {Zheng}}]{Feng_NPJQM2023}%
  \BibitemOpen
  \bibfield  {author} {\bibinfo {author} {\bibfnamefont {X.~Y.}\ \bibnamefont {Feng}}, \bibinfo {author} {\bibfnamefont {Z.}~\bibnamefont {Zhao}}, \bibinfo {author} {\bibfnamefont {J.}~\bibnamefont {Luo}}, \bibinfo {author} {\bibfnamefont {J.}~\bibnamefont {Yang}}, \bibinfo {author} {\bibfnamefont {A.~F.}\ \bibnamefont {Fang}}, \bibinfo {author} {\bibfnamefont {H.~T.}\ \bibnamefont {Yang}}, \bibinfo {author} {\bibfnamefont {H.~J.}\ \bibnamefont {Gao}}, \bibinfo {author} {\bibfnamefont {R.}~\bibnamefont {Zhou}},\ and\ \bibinfo {author} {\bibfnamefont {G.-q.}\ \bibnamefont {Zheng}},\ }\bibfield  {title} {\bibinfo {title} {Commensurate-to-incommensurate transition of charge-density-wave order and a possible quantum critical point in pressurized kagome metal {CsV}$_{3}${Sb}$_{5}$},\ }\href {https://doi.org/10.1038/s41535-023-00555-w} {\bibfield  {journal} {\bibinfo  {journal} {npj Quantum Materials}\ }\textbf {\bibinfo {volume} {8}},\ \bibinfo {pages} {23} (\bibinfo {year} {2023})}\BibitemShut {NoStop}%
\bibitem [{\citenamefont {Mielke}\ \emph {et~al.}(2022)\citenamefont {Mielke}, \citenamefont {Das}, \citenamefont {Yin}, \citenamefont {Liu}, \citenamefont {Gupta}, \citenamefont {Jiang}, \citenamefont {Medarde}, \citenamefont {Wu}, \citenamefont {Lei}, \citenamefont {Chang}, \citenamefont {Dai}, \citenamefont {Si}, \citenamefont {Miao}, \citenamefont {Thomale}, \citenamefont {Neupert}, \citenamefont {Shi}, \citenamefont {Khasanov}, \citenamefont {Hasan}, \citenamefont {Luetkens},\ and\ \citenamefont {Guguchia}}]{Mielke_Nat2022}%
  \BibitemOpen
  \bibfield  {author} {\bibinfo {author} {\bibfnamefont {C.}~\bibnamefont {Mielke}}, \bibinfo {author} {\bibfnamefont {D.}~\bibnamefont {Das}}, \bibinfo {author} {\bibfnamefont {J.-X.}\ \bibnamefont {Yin}}, \bibinfo {author} {\bibfnamefont {H.}~\bibnamefont {Liu}}, \bibinfo {author} {\bibfnamefont {R.}~\bibnamefont {Gupta}}, \bibinfo {author} {\bibfnamefont {Y.-X.}\ \bibnamefont {Jiang}}, \bibinfo {author} {\bibfnamefont {M.}~\bibnamefont {Medarde}}, \bibinfo {author} {\bibfnamefont {X.}~\bibnamefont {Wu}}, \bibinfo {author} {\bibfnamefont {H.~C.}\ \bibnamefont {Lei}}, \bibinfo {author} {\bibfnamefont {J.}~\bibnamefont {Chang}}, \bibinfo {author} {\bibfnamefont {P.}~\bibnamefont {Dai}}, \bibinfo {author} {\bibfnamefont {Q.}~\bibnamefont {Si}}, \bibinfo {author} {\bibfnamefont {H.}~\bibnamefont {Miao}}, \bibinfo {author} {\bibfnamefont {R.}~\bibnamefont {Thomale}}, \bibinfo {author} {\bibfnamefont {T.}~\bibnamefont {Neupert}}, \bibinfo {author} {\bibfnamefont {Y.}~\bibnamefont {Shi}}, \bibinfo {author}
  {\bibfnamefont {R.}~\bibnamefont {Khasanov}}, \bibinfo {author} {\bibfnamefont {M.~Z.}\ \bibnamefont {Hasan}}, \bibinfo {author} {\bibfnamefont {H.}~\bibnamefont {Luetkens}},\ and\ \bibinfo {author} {\bibfnamefont {Z.}~\bibnamefont {Guguchia}},\ }\bibfield  {title} {\bibinfo {title} {Time-reversal symmetry-breaking charge order in a kagome superconductor},\ }\href {https://doi.org/10.1038/s41586-021-04327-z} {\bibfield  {journal} {\bibinfo  {journal} {Nature}\ }\textbf {\bibinfo {volume} {602}},\ \bibinfo {pages} {245} (\bibinfo {year} {2022})}\BibitemShut {NoStop}%
\bibitem [{\citenamefont {Guguchia}\ \emph {et~al.}(2023)\citenamefont {Guguchia}, \citenamefont {Khasanov},\ and\ \citenamefont {Luetkens}}]{Guguchia_NPJ2023}%
  \BibitemOpen
  \bibfield  {author} {\bibinfo {author} {\bibfnamefont {Z.}~\bibnamefont {Guguchia}}, \bibinfo {author} {\bibfnamefont {R.}~\bibnamefont {Khasanov}},\ and\ \bibinfo {author} {\bibfnamefont {H.}~\bibnamefont {Luetkens}},\ }\bibfield  {title} {\bibinfo {title} {Unconventional charge order and superconductivity in kagome-lattice systems as seen by muon-spin rotation},\ }\href {https://doi.org/10.1038/s41535-023-00574-7} {\bibfield  {journal} {\bibinfo  {journal} {npj Quantum Materials}\ }\textbf {\bibinfo {volume} {8}},\ \bibinfo {pages} {41} (\bibinfo {year} {2023})}\BibitemShut {NoStop}%
\bibitem [{\citenamefont {Chen}\ \emph {et~al.}(2021{\natexlab{a}})\citenamefont {Chen}, \citenamefont {Wang}, \citenamefont {Yin}, \citenamefont {Gu}, \citenamefont {Jiang}, \citenamefont {Tu}, \citenamefont {Gong}, \citenamefont {Uwatoko}, \citenamefont {Sun}, \citenamefont {Lei}, \citenamefont {Hu},\ and\ \citenamefont {Cheng}}]{Chen_PRL2021}%
  \BibitemOpen
  \bibfield  {author} {\bibinfo {author} {\bibfnamefont {K.~Y.}\ \bibnamefont {Chen}}, \bibinfo {author} {\bibfnamefont {N.~N.}\ \bibnamefont {Wang}}, \bibinfo {author} {\bibfnamefont {Q.~W.}\ \bibnamefont {Yin}}, \bibinfo {author} {\bibfnamefont {Y.~H.}\ \bibnamefont {Gu}}, \bibinfo {author} {\bibfnamefont {K.}~\bibnamefont {Jiang}}, \bibinfo {author} {\bibfnamefont {Z.~J.}\ \bibnamefont {Tu}}, \bibinfo {author} {\bibfnamefont {C.~S.}\ \bibnamefont {Gong}}, \bibinfo {author} {\bibfnamefont {Y.}~\bibnamefont {Uwatoko}}, \bibinfo {author} {\bibfnamefont {J.~P.}\ \bibnamefont {Sun}}, \bibinfo {author} {\bibfnamefont {H.~C.}\ \bibnamefont {Lei}}, \bibinfo {author} {\bibfnamefont {J.~P.}\ \bibnamefont {Hu}},\ and\ \bibinfo {author} {\bibfnamefont {J.~G.}\ \bibnamefont {Cheng}},\ }\bibfield  {title} {\bibinfo {title} {Double superconducting dome and triple enhancement of ${T}_{c}$ in the kagome superconductor {CsV}$_{3}${Sb}$_{5}$ under high pressure},\ }\href {https://doi.org/10.1103/PhysRevLett.126.247001}
  {\bibfield  {journal} {\bibinfo  {journal} {Physical Review Letters}\ }\textbf {\bibinfo {volume} {126}},\ \bibinfo {pages} {247001} (\bibinfo {year} {2021}{\natexlab{a}})}\BibitemShut {NoStop}%
\bibitem [{\citenamefont {Du}\ \emph {et~al.}(2021{\natexlab{a}})\citenamefont {Du}, \citenamefont {Luo}, \citenamefont {Ortiz}, \citenamefont {Chen}, \citenamefont {Duan}, \citenamefont {Zhang}, \citenamefont {Lu}, \citenamefont {Wilson}, \citenamefont {Song},\ and\ \citenamefont {Yuan}}]{Du_PRB2021}%
  \BibitemOpen
  \bibfield  {author} {\bibinfo {author} {\bibfnamefont {F.}~\bibnamefont {Du}}, \bibinfo {author} {\bibfnamefont {S.}~\bibnamefont {Luo}}, \bibinfo {author} {\bibfnamefont {B.~R.}\ \bibnamefont {Ortiz}}, \bibinfo {author} {\bibfnamefont {Y.}~\bibnamefont {Chen}}, \bibinfo {author} {\bibfnamefont {W.}~\bibnamefont {Duan}}, \bibinfo {author} {\bibfnamefont {D.}~\bibnamefont {Zhang}}, \bibinfo {author} {\bibfnamefont {X.}~\bibnamefont {Lu}}, \bibinfo {author} {\bibfnamefont {S.~D.}\ \bibnamefont {Wilson}}, \bibinfo {author} {\bibfnamefont {Y.}~\bibnamefont {Song}},\ and\ \bibinfo {author} {\bibfnamefont {H.}~\bibnamefont {Yuan}},\ }\bibfield  {title} {\bibinfo {title} {Pressure-induced double superconducting domes and charge instability in the kagome metal {KV}$_{3}${Sb}$_{5}$},\ }\href {https://doi.org/10.1103/PhysRevB.103.L220504} {\bibfield  {journal} {\bibinfo  {journal} {Physical Review B}\ }\textbf {\bibinfo {volume} {103}},\ \bibinfo {pages} {L220504} (\bibinfo {year} {2021}{\natexlab{a}})}\BibitemShut
  {NoStop}%
\bibitem [{\citenamefont {Tan}\ \emph {et~al.}(2021)\citenamefont {Tan}, \citenamefont {Liu}, \citenamefont {Wang},\ and\ \citenamefont {Yan}}]{Tan_PRL2021}%
  \BibitemOpen
  \bibfield  {author} {\bibinfo {author} {\bibfnamefont {H.}~\bibnamefont {Tan}}, \bibinfo {author} {\bibfnamefont {Y.}~\bibnamefont {Liu}}, \bibinfo {author} {\bibfnamefont {Z.}~\bibnamefont {Wang}},\ and\ \bibinfo {author} {\bibfnamefont {B.}~\bibnamefont {Yan}},\ }\bibfield  {title} {\bibinfo {title} {Charge density waves and electronic properties of superconducting kagome metals},\ }\href {https://doi.org/10.1103/PhysRevLett.127.046401} {\bibfield  {journal} {\bibinfo  {journal} {Physical Review Letters}\ }\textbf {\bibinfo {volume} {127}},\ \bibinfo {pages} {046401} (\bibinfo {year} {2021})}\BibitemShut {NoStop}%
\bibitem [{\citenamefont {Kang}\ \emph {et~al.}(2022)\citenamefont {Kang}, \citenamefont {Fang}, \citenamefont {Kim}, \citenamefont {Ortiz}, \citenamefont {Ryu}, \citenamefont {Kim}, \citenamefont {Yoo}, \citenamefont {Sangiovanni}, \citenamefont {Di~Sante}, \citenamefont {Park}, \citenamefont {Jozwiak}, \citenamefont {Bostwick}, \citenamefont {Rotenberg}, \citenamefont {Kaxiras}, \citenamefont {Wilson}, \citenamefont {Park},\ and\ \citenamefont {Comin}}]{Kang_NaturePhysics2022}%
  \BibitemOpen
  \bibfield  {author} {\bibinfo {author} {\bibfnamefont {M.}~\bibnamefont {Kang}}, \bibinfo {author} {\bibfnamefont {S.}~\bibnamefont {Fang}}, \bibinfo {author} {\bibfnamefont {J.-K.}\ \bibnamefont {Kim}}, \bibinfo {author} {\bibfnamefont {B.~R.}\ \bibnamefont {Ortiz}}, \bibinfo {author} {\bibfnamefont {S.~H.}\ \bibnamefont {Ryu}}, \bibinfo {author} {\bibfnamefont {J.}~\bibnamefont {Kim}}, \bibinfo {author} {\bibfnamefont {J.}~\bibnamefont {Yoo}}, \bibinfo {author} {\bibfnamefont {G.}~\bibnamefont {Sangiovanni}}, \bibinfo {author} {\bibfnamefont {D.}~\bibnamefont {Di~Sante}}, \bibinfo {author} {\bibfnamefont {B.-G.}\ \bibnamefont {Park}}, \bibinfo {author} {\bibfnamefont {C.}~\bibnamefont {Jozwiak}}, \bibinfo {author} {\bibfnamefont {A.}~\bibnamefont {Bostwick}}, \bibinfo {author} {\bibfnamefont {E.}~\bibnamefont {Rotenberg}}, \bibinfo {author} {\bibfnamefont {E.}~\bibnamefont {Kaxiras}}, \bibinfo {author} {\bibfnamefont {S.~D.}\ \bibnamefont {Wilson}}, \bibinfo {author} {\bibfnamefont {J.-H.}\ \bibnamefont
  {Park}},\ and\ \bibinfo {author} {\bibfnamefont {R.}~\bibnamefont {Comin}},\ }\bibfield  {title} {\bibinfo {title} {Twofold van {H}ove singularity and origin of charge order in topological kagome superconductor {CsV}$_{3}${Sb}$_{5}$},\ }\href {https://doi.org/10.1038/s41567-021-01451-5} {\bibfield  {journal} {\bibinfo  {journal} {Nature Physics}\ }\textbf {\bibinfo {volume} {18}},\ \bibinfo {pages} {301} (\bibinfo {year} {2022})}\BibitemShut {NoStop}%
\bibitem [{SOM()}]{SOM}%
  \BibitemOpen
  \href@noop {} {}\bibinfo {note} {See Supplemental Material at XXX for information on the samples and experimental methods as well as on the detailed structural analysis, which includes also references~\cite{Crysalis, ShelXT,Jana2006,Stahl_PRB22,Zhang_PRB2021,Zhang_PRB2021,Yu_PRL2021,Tsirlin_PRB2023,Koepernik_PRB1999}.}\BibitemShut {Stop}%
\bibitem [{\citenamefont {Ortiz}\ \emph {et~al.}(2019)\citenamefont {Ortiz}, \citenamefont {Gomes}, \citenamefont {Morey}, \citenamefont {Winiarski}, \citenamefont {Bordelon}, \citenamefont {Mangum}, \citenamefont {Oswald}, \citenamefont {Rodriguez-Rivera}, \citenamefont {Neilson}, \citenamefont {Wilson}, \citenamefont {Ertekin}, \citenamefont {McQueen},\ and\ \citenamefont {Toberer}}]{Ortiz_PRM2019}%
  \BibitemOpen
  \bibfield  {author} {\bibinfo {author} {\bibfnamefont {B.~R.}\ \bibnamefont {Ortiz}}, \bibinfo {author} {\bibfnamefont {L.~C.}\ \bibnamefont {Gomes}}, \bibinfo {author} {\bibfnamefont {J.~R.}\ \bibnamefont {Morey}}, \bibinfo {author} {\bibfnamefont {M.}~\bibnamefont {Winiarski}}, \bibinfo {author} {\bibfnamefont {M.}~\bibnamefont {Bordelon}}, \bibinfo {author} {\bibfnamefont {J.~S.}\ \bibnamefont {Mangum}}, \bibinfo {author} {\bibfnamefont {I.~W.~H.}\ \bibnamefont {Oswald}}, \bibinfo {author} {\bibfnamefont {J.~A.}\ \bibnamefont {Rodriguez-Rivera}}, \bibinfo {author} {\bibfnamefont {J.~R.}\ \bibnamefont {Neilson}}, \bibinfo {author} {\bibfnamefont {S.~D.}\ \bibnamefont {Wilson}}, \bibinfo {author} {\bibfnamefont {E.}~\bibnamefont {Ertekin}}, \bibinfo {author} {\bibfnamefont {T.~M.}\ \bibnamefont {McQueen}},\ and\ \bibinfo {author} {\bibfnamefont {E.~S.}\ \bibnamefont {Toberer}},\ }\bibfield  {title} {\bibinfo {title} {New kagome prototype materials: discovery of {KV}$_3${Sb}$_5$, {RbV}$_3${Sb}$_5$, and
  {CsV}$_3${Sb}$_5$},\ }\href {https://doi.org/10.1103/PhysRevMaterials.3.094407} {\bibfield  {journal} {\bibinfo  {journal} {Phys. Rev. Mater.}\ }\textbf {\bibinfo {volume} {3}},\ \bibinfo {pages} {094407} (\bibinfo {year} {2019})}\BibitemShut {NoStop}%
\bibitem [{\citenamefont {Xiao}\ \emph {et~al.}(2023)\citenamefont {Xiao}, \citenamefont {Lin}, \citenamefont {Li}, \citenamefont {Zheng}, \citenamefont {Francoual}, \citenamefont {Plueckthun}, \citenamefont {Xia}, \citenamefont {Qiu}, \citenamefont {Zhang}, \citenamefont {Guo}, \citenamefont {Feng},\ and\ \citenamefont {Peng}}]{Xiao_PRR2023}%
  \BibitemOpen
  \bibfield  {author} {\bibinfo {author} {\bibfnamefont {Q.}~\bibnamefont {Xiao}}, \bibinfo {author} {\bibfnamefont {Y.}~\bibnamefont {Lin}}, \bibinfo {author} {\bibfnamefont {Q.}~\bibnamefont {Li}}, \bibinfo {author} {\bibfnamefont {X.}~\bibnamefont {Zheng}}, \bibinfo {author} {\bibfnamefont {S.}~\bibnamefont {Francoual}}, \bibinfo {author} {\bibfnamefont {C.}~\bibnamefont {Plueckthun}}, \bibinfo {author} {\bibfnamefont {W.}~\bibnamefont {Xia}}, \bibinfo {author} {\bibfnamefont {Q.}~\bibnamefont {Qiu}}, \bibinfo {author} {\bibfnamefont {S.}~\bibnamefont {Zhang}}, \bibinfo {author} {\bibfnamefont {Y.}~\bibnamefont {Guo}}, \bibinfo {author} {\bibfnamefont {J.}~\bibnamefont {Feng}},\ and\ \bibinfo {author} {\bibfnamefont {Y.}~\bibnamefont {Peng}},\ }\bibfield  {title} {\bibinfo {title} {Coexistence of multiple stacking charge density waves in kagome superconductor {CsV}$_3${Sb}$_5$},\ }\href {https://doi.org/10.1103/PhysRevResearch.5.L012032} {\bibfield  {journal} {\bibinfo  {journal} {Physical Review
  Research}\ }\textbf {\bibinfo {volume} {5}},\ \bibinfo {pages} {L012032} (\bibinfo {year} {2023})}\BibitemShut {NoStop}%
\bibitem [{\citenamefont {Ortiz}\ \emph {et~al.}(2021)\citenamefont {Ortiz}, \citenamefont {Teicher}, \citenamefont {Kautzsch}, \citenamefont {Sarte}, \citenamefont {Ratcliff}, \citenamefont {Harter}, \citenamefont {Ruff}, \citenamefont {Seshadri},\ and\ \citenamefont {Wilson}}]{Ortiz_PRX2021}%
  \BibitemOpen
  \bibfield  {author} {\bibinfo {author} {\bibfnamefont {B.~R.}\ \bibnamefont {Ortiz}}, \bibinfo {author} {\bibfnamefont {S.~M.~L.}\ \bibnamefont {Teicher}}, \bibinfo {author} {\bibfnamefont {L.}~\bibnamefont {Kautzsch}}, \bibinfo {author} {\bibfnamefont {P.~M.}\ \bibnamefont {Sarte}}, \bibinfo {author} {\bibfnamefont {N.}~\bibnamefont {Ratcliff}}, \bibinfo {author} {\bibfnamefont {J.}~\bibnamefont {Harter}}, \bibinfo {author} {\bibfnamefont {J.~P.~C.}\ \bibnamefont {Ruff}}, \bibinfo {author} {\bibfnamefont {R.}~\bibnamefont {Seshadri}},\ and\ \bibinfo {author} {\bibfnamefont {S.~D.}\ \bibnamefont {Wilson}},\ }\bibfield  {title} {\bibinfo {title} {Fermi surface mapping and the nature of charge-density-wave order in the kagome superconductor {CsV}$_{3}${Sb}$_{5}$},\ }\href {https://doi.org/10.1103/PhysRevX.11.041030} {\bibfield  {journal} {\bibinfo  {journal} {Phys. Rev. X}\ }\textbf {\bibinfo {volume} {11}},\ \bibinfo {pages} {041030} (\bibinfo {year} {2021})}\BibitemShut {NoStop}%
\bibitem [{\citenamefont {Stahl}\ \emph {et~al.}(2022)\citenamefont {Stahl}, \citenamefont {Chen}, \citenamefont {Ritschel}, \citenamefont {Shekhar}, \citenamefont {Sadrollahi}, \citenamefont {Rahn}, \citenamefont {Ivashko}, \citenamefont {Zimmermann}, \citenamefont {Felser},\ and\ \citenamefont {Geck}}]{Stahl_PRB22}%
  \BibitemOpen
  \bibfield  {author} {\bibinfo {author} {\bibfnamefont {Q.}~\bibnamefont {Stahl}}, \bibinfo {author} {\bibfnamefont {D.}~\bibnamefont {Chen}}, \bibinfo {author} {\bibfnamefont {T.}~\bibnamefont {Ritschel}}, \bibinfo {author} {\bibfnamefont {C.}~\bibnamefont {Shekhar}}, \bibinfo {author} {\bibfnamefont {E.}~\bibnamefont {Sadrollahi}}, \bibinfo {author} {\bibfnamefont {M.~C.}\ \bibnamefont {Rahn}}, \bibinfo {author} {\bibfnamefont {O.}~\bibnamefont {Ivashko}}, \bibinfo {author} {\bibfnamefont {M.~v.}\ \bibnamefont {Zimmermann}}, \bibinfo {author} {\bibfnamefont {C.}~\bibnamefont {Felser}},\ and\ \bibinfo {author} {\bibfnamefont {J.}~\bibnamefont {Geck}},\ }\bibfield  {title} {\bibinfo {title} {Temperature-driven reorganization of electronic order in {CsV}$_{3}${Sb}$_{5}$},\ }\href {https://doi.org/10.1103/PhysRevB.105.195136} {\bibfield  {journal} {\bibinfo  {journal} {Phys. Rev. B}\ }\textbf {\bibinfo {volume} {105}},\ \bibinfo {pages} {195136} (\bibinfo {year} {2022})}\BibitemShut {NoStop}%
\bibitem [{\citenamefont {Chen}\ \emph {et~al.}(2021{\natexlab{b}})\citenamefont {Chen}, \citenamefont {Wang}, \citenamefont {Yin}, \citenamefont {Gu}, \citenamefont {Jiang}, \citenamefont {Tu}, \citenamefont {Gong}, \citenamefont {Uwatoko}, \citenamefont {Sun}, \citenamefont {Lei}, \citenamefont {Hu},\ and\ \citenamefont {Cheng}}]{Chen_PRL21}%
  \BibitemOpen
  \bibfield  {author} {\bibinfo {author} {\bibfnamefont {K.~Y.}\ \bibnamefont {Chen}}, \bibinfo {author} {\bibfnamefont {N.~N.}\ \bibnamefont {Wang}}, \bibinfo {author} {\bibfnamefont {Q.~W.}\ \bibnamefont {Yin}}, \bibinfo {author} {\bibfnamefont {Y.~H.}\ \bibnamefont {Gu}}, \bibinfo {author} {\bibfnamefont {K.}~\bibnamefont {Jiang}}, \bibinfo {author} {\bibfnamefont {Z.~J.}\ \bibnamefont {Tu}}, \bibinfo {author} {\bibfnamefont {C.~S.}\ \bibnamefont {Gong}}, \bibinfo {author} {\bibfnamefont {Y.}~\bibnamefont {Uwatoko}}, \bibinfo {author} {\bibfnamefont {J.~P.}\ \bibnamefont {Sun}}, \bibinfo {author} {\bibfnamefont {H.~C.}\ \bibnamefont {Lei}}, \bibinfo {author} {\bibfnamefont {J.~P.}\ \bibnamefont {Hu}},\ and\ \bibinfo {author} {\bibfnamefont {J.-G.}\ \bibnamefont {Cheng}},\ }\bibfield  {title} {\bibinfo {title} {Double superconducting dome and triple enhancement of ${T}_{c}$ in the kagome superconductor {CsV}$_{3}${Sb}$_{5}$ under high pressure},\ }\href {https://doi.org/10.1103/PhysRevLett.126.247001}
  {\bibfield  {journal} {\bibinfo  {journal} {Phys. Rev. Lett.}\ }\textbf {\bibinfo {volume} {126}},\ \bibinfo {pages} {247001} (\bibinfo {year} {2021}{\natexlab{b}})}\BibitemShut {NoStop}%
\bibitem [{\citenamefont {Du}\ \emph {et~al.}(2021{\natexlab{b}})\citenamefont {Du}, \citenamefont {Luo}, \citenamefont {Ortiz}, \citenamefont {Chen}, \citenamefont {Duan}, \citenamefont {Zhang}, \citenamefont {Lu}, \citenamefont {Wilson}, \citenamefont {Song},\ and\ \citenamefont {Yuan}}]{Feng_PRB21}%
  \BibitemOpen
  \bibfield  {author} {\bibinfo {author} {\bibfnamefont {F.}~\bibnamefont {Du}}, \bibinfo {author} {\bibfnamefont {S.}~\bibnamefont {Luo}}, \bibinfo {author} {\bibfnamefont {B.~R.}\ \bibnamefont {Ortiz}}, \bibinfo {author} {\bibfnamefont {Y.}~\bibnamefont {Chen}}, \bibinfo {author} {\bibfnamefont {W.}~\bibnamefont {Duan}}, \bibinfo {author} {\bibfnamefont {D.}~\bibnamefont {Zhang}}, \bibinfo {author} {\bibfnamefont {X.}~\bibnamefont {Lu}}, \bibinfo {author} {\bibfnamefont {S.~D.}\ \bibnamefont {Wilson}}, \bibinfo {author} {\bibfnamefont {Y.}~\bibnamefont {Song}},\ and\ \bibinfo {author} {\bibfnamefont {H.}~\bibnamefont {Yuan}},\ }\bibfield  {title} {\bibinfo {title} {Pressure-induced double superconducting domes and charge instability in the kagome metal {KV}$_{3}${Sb}$_{5}$},\ }\href {https://doi.org/10.1103/PhysRevB.103.L220504} {\bibfield  {journal} {\bibinfo  {journal} {Phys. Rev. B}\ }\textbf {\bibinfo {volume} {103}},\ \bibinfo {pages} {L220504} (\bibinfo {year} {2021}{\natexlab{b}})}\BibitemShut
  {NoStop}%
\bibitem [{\citenamefont {Hossain}\ \emph {et~al.}(2013)\citenamefont {Hossain}, \citenamefont {Zegkinoglou}, \citenamefont {Chuang}, \citenamefont {Geck}, \citenamefont {Bohnenbuck}, \citenamefont {Gonzalez}, \citenamefont {Wu}, \citenamefont {Sch{\"u}{\ss}ler-Langeheine}, \citenamefont {Hawthorn}, \citenamefont {Denlinger}, \citenamefont {Mathieu}, \citenamefont {Tokura}, \citenamefont {Satow}, \citenamefont {Takagi}, \citenamefont {Yoshida}, \citenamefont {Hussain}, \citenamefont {Keimer}, \citenamefont {Sawatzky},\ and\ \citenamefont {Damascelli}}]{Hossain:2013a}%
  \BibitemOpen
  \bibfield  {author} {\bibinfo {author} {\bibfnamefont {M.~A.}\ \bibnamefont {Hossain}}, \bibinfo {author} {\bibfnamefont {I.}~\bibnamefont {Zegkinoglou}}, \bibinfo {author} {\bibfnamefont {Y.~D.}\ \bibnamefont {Chuang}}, \bibinfo {author} {\bibfnamefont {J.}~\bibnamefont {Geck}}, \bibinfo {author} {\bibfnamefont {B.}~\bibnamefont {Bohnenbuck}}, \bibinfo {author} {\bibfnamefont {A.~G.~C.}\ \bibnamefont {Gonzalez}}, \bibinfo {author} {\bibfnamefont {H.~H.}\ \bibnamefont {Wu}}, \bibinfo {author} {\bibfnamefont {C.}~\bibnamefont {Sch{\"u}{\ss}ler-Langeheine}}, \bibinfo {author} {\bibfnamefont {D.~G.}\ \bibnamefont {Hawthorn}}, \bibinfo {author} {\bibfnamefont {J.~D.}\ \bibnamefont {Denlinger}}, \bibinfo {author} {\bibfnamefont {R.}~\bibnamefont {Mathieu}}, \bibinfo {author} {\bibfnamefont {Y.}~\bibnamefont {Tokura}}, \bibinfo {author} {\bibfnamefont {S.}~\bibnamefont {Satow}}, \bibinfo {author} {\bibfnamefont {H.}~\bibnamefont {Takagi}}, \bibinfo {author} {\bibfnamefont {Y.}~\bibnamefont {Yoshida}}, \bibinfo
  {author} {\bibfnamefont {Z.}~\bibnamefont {Hussain}}, \bibinfo {author} {\bibfnamefont {B.}~\bibnamefont {Keimer}}, \bibinfo {author} {\bibfnamefont {G.~A.}\ \bibnamefont {Sawatzky}},\ and\ \bibinfo {author} {\bibfnamefont {A.}~\bibnamefont {Damascelli}},\ }\bibfield  {title} {\bibinfo {title} {Electronic superlattice revealed by resonant scattering from random impurities in {Sr}$_{3}${Ru}$_{2}${O}$_{7}$},\ }\href {https://doi.org/https://doi.org/10.1038/srep02299} {\bibfield  {journal} {\bibinfo  {journal} {Scientific Reports}\ }\textbf {\bibinfo {volume} {3}},\ \bibinfo {pages} {2299} (\bibinfo {year} {2013})}\BibitemShut {NoStop}%
\bibitem [{\citenamefont {Kautzsch}\ \emph {et~al.}(2023)\citenamefont {Kautzsch}, \citenamefont {Oey}, \citenamefont {Li}, \citenamefont {Ren}, \citenamefont {Ortiz}, \citenamefont {Pokharel}, \citenamefont {Seshadri}, \citenamefont {Ruff}, \citenamefont {Kongruengkit}, \citenamefont {Harter}, \citenamefont {Wang}, \citenamefont {Zeljkovic},\ and\ \citenamefont {Wilson}}]{Kautzsch_npjQM23}%
  \BibitemOpen
  \bibfield  {author} {\bibinfo {author} {\bibfnamefont {L.}~\bibnamefont {Kautzsch}}, \bibinfo {author} {\bibfnamefont {Y.~M.}\ \bibnamefont {Oey}}, \bibinfo {author} {\bibfnamefont {H.}~\bibnamefont {Li}}, \bibinfo {author} {\bibfnamefont {Z.}~\bibnamefont {Ren}}, \bibinfo {author} {\bibfnamefont {B.~R.}\ \bibnamefont {Ortiz}}, \bibinfo {author} {\bibfnamefont {G.}~\bibnamefont {Pokharel}}, \bibinfo {author} {\bibfnamefont {R.}~\bibnamefont {Seshadri}}, \bibinfo {author} {\bibfnamefont {J.}~\bibnamefont {Ruff}}, \bibinfo {author} {\bibfnamefont {T.}~\bibnamefont {Kongruengkit}}, \bibinfo {author} {\bibfnamefont {J.~W.}\ \bibnamefont {Harter}}, \bibinfo {author} {\bibfnamefont {Z.}~\bibnamefont {Wang}}, \bibinfo {author} {\bibfnamefont {I.}~\bibnamefont {Zeljkovic}},\ and\ \bibinfo {author} {\bibfnamefont {S.~D.}\ \bibnamefont {Wilson}},\ }\bibfield  {title} {\bibinfo {title} {Incommensurate charge-stripe correlations in the kagome superconductor {CsV}$_{3}${Sb}$_{5-x}${Sn}$_{x}$},\ }\href
  {https://doi.org/10.1038/s41535-023-00570-x} {\bibfield  {journal} {\bibinfo  {journal} {npj Quantum Materials}\ }\textbf {\bibinfo {volume} {8}},\ \bibinfo {pages} {37} (\bibinfo {year} {2023})}\BibitemShut {NoStop}%
\bibitem [{\citenamefont {Wenzel}\ \emph {et~al.}(2023)\citenamefont {Wenzel}, \citenamefont {Tsirlin}, \citenamefont {Capitani}, \citenamefont {Chan}, \citenamefont {Ortiz}, \citenamefont {Wilson}, \citenamefont {Dressel},\ and\ \citenamefont {Uykur}}]{Wenzel:2023a}%
  \BibitemOpen
  \bibfield  {author} {\bibinfo {author} {\bibfnamefont {M.}~\bibnamefont {Wenzel}}, \bibinfo {author} {\bibfnamefont {A.~A.}\ \bibnamefont {Tsirlin}}, \bibinfo {author} {\bibfnamefont {F.}~\bibnamefont {Capitani}}, \bibinfo {author} {\bibfnamefont {Y.~T.}\ \bibnamefont {Chan}}, \bibinfo {author} {\bibfnamefont {B.~R.}\ \bibnamefont {Ortiz}}, \bibinfo {author} {\bibfnamefont {S.~D.}\ \bibnamefont {Wilson}}, \bibinfo {author} {\bibfnamefont {M.}~\bibnamefont {Dressel}},\ and\ \bibinfo {author} {\bibfnamefont {E.}~\bibnamefont {Uykur}},\ }\bibfield  {title} {\bibinfo {title} {Pressure evolution of electron dynamics in the superconducting kagome metal {CsV}$_{3}${Sb}$_{5}$},\ }\href {https://doi.org/https://doi.org/10.1038/s41535-023-00577-4} {\bibfield  {journal} {\bibinfo  {journal} {npj Quantum Materials}\ }\textbf {\bibinfo {volume} {8}},\ \bibinfo {pages} {45} (\bibinfo {year} {2023})}\BibitemShut {NoStop}%
\bibitem [{\citenamefont {Tsirlin}\ \emph {et~al.}(2022)\citenamefont {Tsirlin}, \citenamefont {Fertey}, \citenamefont {Ortiz}, \citenamefont {Klis}, \citenamefont {Merkl}, \citenamefont {Dressel}, \citenamefont {Wilson},\ and\ \citenamefont {Uykur}}]{Tsirlin_SciPost22}%
  \BibitemOpen
  \bibfield  {author} {\bibinfo {author} {\bibfnamefont {A.~A.}\ \bibnamefont {Tsirlin}}, \bibinfo {author} {\bibfnamefont {P.}~\bibnamefont {Fertey}}, \bibinfo {author} {\bibfnamefont {B.~R.}\ \bibnamefont {Ortiz}}, \bibinfo {author} {\bibfnamefont {B.}~\bibnamefont {Klis}}, \bibinfo {author} {\bibfnamefont {V.}~\bibnamefont {Merkl}}, \bibinfo {author} {\bibfnamefont {M.}~\bibnamefont {Dressel}}, \bibinfo {author} {\bibfnamefont {S.~D.}\ \bibnamefont {Wilson}},\ and\ \bibinfo {author} {\bibfnamefont {E.}~\bibnamefont {Uykur}},\ }\bibfield  {title} {\bibinfo {title} {Role of sb in the superconducting kagome metal {CsV}$_3${Sb}$_5$ revealed by its anisotropic compression},\ }\href {https://doi.org/10.21468/SciPostPhys.12.2.049} {\bibfield  {journal} {\bibinfo  {journal} {SciPost Phys.}\ }\textbf {\bibinfo {volume} {12}},\ \bibinfo {pages} {049} (\bibinfo {year} {2022})}\BibitemShut {NoStop}%
\bibitem [{\citenamefont {Kaboudvand}\ \emph {et~al.}(2022)\citenamefont {Kaboudvand}, \citenamefont {Teicher}, \citenamefont {Wilson}, \citenamefont {Seshadri},\ and\ \citenamefont {Johannes}}]{Kaboudvand:2022aa}%
  \BibitemOpen
  \bibfield  {author} {\bibinfo {author} {\bibfnamefont {F.}~\bibnamefont {Kaboudvand}}, \bibinfo {author} {\bibfnamefont {S.~M.~L.}\ \bibnamefont {Teicher}}, \bibinfo {author} {\bibfnamefont {S.~D.}\ \bibnamefont {Wilson}}, \bibinfo {author} {\bibfnamefont {R.}~\bibnamefont {Seshadri}},\ and\ \bibinfo {author} {\bibfnamefont {M.~D.}\ \bibnamefont {Johannes}},\ }\bibfield  {title} {\bibinfo {title} {Fermi surface nesting and the lindhard response function in the kagome superconductor {CsV}$_3${Sb}$_5$},\ }\href {https://doi.org/10.1063/5.0081081} {\bibfield  {journal} {\bibinfo  {journal} {Applied Physics Letters}\ }\textbf {\bibinfo {volume} {120}},\ \bibinfo {pages} {111901} (\bibinfo {year} {2022})}\BibitemShut {NoStop}%
\bibitem [{\citenamefont {Xie}\ \emph {et~al.}(2022)\citenamefont {Xie}, \citenamefont {Li}, \citenamefont {Bourges}, \citenamefont {Ivanov}, \citenamefont {Ye}, \citenamefont {Yin}, \citenamefont {Hasan}, \citenamefont {Luo}, \citenamefont {Yao}, \citenamefont {Wang}, \citenamefont {Xu},\ and\ \citenamefont {Dai}}]{Xie:2022a}%
  \BibitemOpen
  \bibfield  {author} {\bibinfo {author} {\bibfnamefont {Y.}~\bibnamefont {Xie}}, \bibinfo {author} {\bibfnamefont {Y.}~\bibnamefont {Li}}, \bibinfo {author} {\bibfnamefont {P.}~\bibnamefont {Bourges}}, \bibinfo {author} {\bibfnamefont {A.}~\bibnamefont {Ivanov}}, \bibinfo {author} {\bibfnamefont {Z.}~\bibnamefont {Ye}}, \bibinfo {author} {\bibfnamefont {J.-X.}\ \bibnamefont {Yin}}, \bibinfo {author} {\bibfnamefont {M.~Z.}\ \bibnamefont {Hasan}}, \bibinfo {author} {\bibfnamefont {A.}~\bibnamefont {Luo}}, \bibinfo {author} {\bibfnamefont {Y.}~\bibnamefont {Yao}}, \bibinfo {author} {\bibfnamefont {Z.}~\bibnamefont {Wang}}, \bibinfo {author} {\bibfnamefont {G.}~\bibnamefont {Xu}},\ and\ \bibinfo {author} {\bibfnamefont {P.}~\bibnamefont {Dai}},\ }\bibfield  {title} {\bibinfo {title} {Electron-phonon coupling in the charge density wave state of {CsV}$_3${Sb}$_5$},\ }\href {https://doi.org/https://doi.org/10.1103/PhysRevB.105.L140501} {\bibfield  {journal} {\bibinfo  {journal} {Phys. Rev. B}\ }\textbf {\bibinfo
  {volume} {105}},\ \bibinfo {pages} {L140501} (\bibinfo {year} {2022})}\BibitemShut {NoStop}%
\bibitem [{\citenamefont {Ye}\ \emph {et~al.}(2022)\citenamefont {Ye}, \citenamefont {Luo}, \citenamefont {Yin}, \citenamefont {Hasan},\ and\ \citenamefont {Xu}}]{Ye:2022a}%
  \BibitemOpen
  \bibfield  {author} {\bibinfo {author} {\bibfnamefont {Z.}~\bibnamefont {Ye}}, \bibinfo {author} {\bibfnamefont {A.}~\bibnamefont {Luo}}, \bibinfo {author} {\bibfnamefont {J.-X.}\ \bibnamefont {Yin}}, \bibinfo {author} {\bibfnamefont {M.~Z.}\ \bibnamefont {Hasan}},\ and\ \bibinfo {author} {\bibfnamefont {G.}~\bibnamefont {Xu}},\ }\bibfield  {title} {\bibinfo {title} {Structural instability and charge modulations in the kagome superconductor {AV}$_{3}${Sb}$_{5}$},\ }\href {https://doi.org/https://doi.org/10.1103/PhysRevB.105.245121} {\bibfield  {journal} {\bibinfo  {journal} {Phys. Rev. B}\ }\textbf {\bibinfo {volume} {105}},\ \bibinfo {pages} {245121} (\bibinfo {year} {2022})}\BibitemShut {NoStop}%
\bibitem [{\citenamefont {Gutierrez-Amigo}\ \emph {et~al.}(2024)\citenamefont {Gutierrez-Amigo}, \citenamefont {DangiÄ}, \citenamefont {Guo}, \citenamefont {Felser}, \citenamefont {Moll}, \citenamefont {Vergniory},\ and\ \citenamefont {Errea}}]{gutierrez_2023}%
  \BibitemOpen
  \bibfield  {author} {\bibinfo {author} {\bibfnamefont {M.}~\bibnamefont {Gutierrez-Amigo}}, \bibinfo {author} {\bibfnamefont {Ã.}~\bibnamefont {Dangi\'c}}, \bibinfo {author} {\bibfnamefont {C.}~\bibnamefont {Guo}}, \bibinfo {author} {\bibfnamefont {C.}~\bibnamefont {Felser}}, \bibinfo {author} {\bibfnamefont {P.~J.~W.}\ \bibnamefont {Moll}}, \bibinfo {author} {\bibfnamefont {M.~G.}\ \bibnamefont {Vergniory}},\ and\ \bibinfo {author} {\bibfnamefont {I.}~\bibnamefont {Errea}},\ }\bibfield  {title} {\bibinfo {title} {Phonon collapse and anharmonic melting of the 3d charge-density wave in kagome metals},\ }\href {https://doi.org/10.1038/s43246-024-00676-0} {\bibfield  {journal} {\bibinfo  {journal} {Communications Materials}\ }\textbf {\bibinfo {volume} {5}},\ \bibinfo {pages} {234} (\bibinfo {year} {2024})}\BibitemShut {NoStop}%
\bibitem [{Cry()}]{Crysalis}%
  \BibitemOpen
  \href@noop {} {\bibinfo {title} {Crysalis$^{\mathrm{pro}}$ single crystal x-ray diffraction data collection and processing software}}\BibitemShut {NoStop}%
\bibitem [{\citenamefont {Sheldrick}(2015)}]{ShelXT}%
  \BibitemOpen
  \bibfield  {author} {\bibinfo {author} {\bibfnamefont {G.~M.}\ \bibnamefont {Sheldrick}},\ }\bibfield  {title} {\bibinfo {title} {{{\it SHELXT} {--} Integrated space-group and crystal-structure determination}},\ }\href {https://doi.org/10.1107/S2053273314026370} {\bibfield  {journal} {\bibinfo  {journal} {Acta Crystallographica Section A}\ }\textbf {\bibinfo {volume} {71}},\ \bibinfo {pages} {3} (\bibinfo {year} {2015})}\BibitemShut {NoStop}%
\bibitem [{\citenamefont {Pet\v{r}\'{i}\v{c}ek}\ \emph {et~al.}(2014)\citenamefont {Pet\v{r}\'{i}\v{c}ek}, \citenamefont {Du\v{s}ek},\ and\ \citenamefont {Palatinus}}]{Jana2006}%
  \BibitemOpen
  \bibfield  {author} {\bibinfo {author} {\bibfnamefont {V.}~\bibnamefont {Pet\v{r}\'{i}\v{c}ek}}, \bibinfo {author} {\bibfnamefont {M.}~\bibnamefont {Du\v{s}ek}},\ and\ \bibinfo {author} {\bibfnamefont {L.}~\bibnamefont {Palatinus}},\ }\bibfield  {title} {\bibinfo {title} {{C}rystallographic {C}omputing {S}ystem {JANA}2006: {G}eneral features},\ }\href {https://doi.org/doi:10.1515/zkri-2014-1737} {\bibfield  {journal} {\bibinfo  {journal} {Zeitschrift fÃŒr Kristallographie - Crystalline Materials}\ }\textbf {\bibinfo {volume} {229}},\ \bibinfo {pages} {345} (\bibinfo {year} {2014})}\BibitemShut {NoStop}%
\bibitem [{\citenamefont {Zhang}\ \emph {et~al.}(2021)\citenamefont {Zhang}, \citenamefont {Chen}, \citenamefont {Zhou}, \citenamefont {Yuan}, \citenamefont {Wang}, \citenamefont {Wang}, \citenamefont {Yang}, \citenamefont {An}, \citenamefont {Zhang}, \citenamefont {Zhu}, \citenamefont {Zhou}, \citenamefont {Chen}, \citenamefont {Zhou},\ and\ \citenamefont {Yang}}]{Zhang_PRB2021}%
  \BibitemOpen
  \bibfield  {author} {\bibinfo {author} {\bibfnamefont {Z.}~\bibnamefont {Zhang}}, \bibinfo {author} {\bibfnamefont {Z.}~\bibnamefont {Chen}}, \bibinfo {author} {\bibfnamefont {Y.}~\bibnamefont {Zhou}}, \bibinfo {author} {\bibfnamefont {Y.}~\bibnamefont {Yuan}}, \bibinfo {author} {\bibfnamefont {S.}~\bibnamefont {Wang}}, \bibinfo {author} {\bibfnamefont {J.}~\bibnamefont {Wang}}, \bibinfo {author} {\bibfnamefont {H.}~\bibnamefont {Yang}}, \bibinfo {author} {\bibfnamefont {C.}~\bibnamefont {An}}, \bibinfo {author} {\bibfnamefont {L.}~\bibnamefont {Zhang}}, \bibinfo {author} {\bibfnamefont {X.}~\bibnamefont {Zhu}}, \bibinfo {author} {\bibfnamefont {Y.}~\bibnamefont {Zhou}}, \bibinfo {author} {\bibfnamefont {X.}~\bibnamefont {Chen}}, \bibinfo {author} {\bibfnamefont {J.}~\bibnamefont {Zhou}},\ and\ \bibinfo {author} {\bibfnamefont {Z.}~\bibnamefont {Yang}},\ }\bibfield  {title} {\bibinfo {title} {Pressure-induced reemergence of superconductivity in the topological kagome metal {CsV}$_{3}${Sb}$_{5}$},\ }\href
  {https://doi.org/10.1103/PhysRevB.103.224513} {\bibfield  {journal} {\bibinfo  {journal} {Phys. Rev. B}\ }\textbf {\bibinfo {volume} {103}},\ \bibinfo {pages} {224513} (\bibinfo {year} {2021})}\BibitemShut {NoStop}%
\bibitem [{\citenamefont {Yu}\ \emph {et~al.}(2022)\citenamefont {Yu}, \citenamefont {Zhu}, \citenamefont {Wen}, \citenamefont {Gui}, \citenamefont {Li}, \citenamefont {Han}, \citenamefont {Wu}, \citenamefont {Wang}, \citenamefont {Xiang}, \citenamefont {Qiao}, \citenamefont {Ying},\ and\ \citenamefont {Chen}}]{Yu_PRL2021}%
  \BibitemOpen
  \bibfield  {author} {\bibinfo {author} {\bibfnamefont {F.}~\bibnamefont {Yu}}, \bibinfo {author} {\bibfnamefont {X.}~\bibnamefont {Zhu}}, \bibinfo {author} {\bibfnamefont {X.}~\bibnamefont {Wen}}, \bibinfo {author} {\bibfnamefont {Z.}~\bibnamefont {Gui}}, \bibinfo {author} {\bibfnamefont {Z.}~\bibnamefont {Li}}, \bibinfo {author} {\bibfnamefont {Y.}~\bibnamefont {Han}}, \bibinfo {author} {\bibfnamefont {T.}~\bibnamefont {Wu}}, \bibinfo {author} {\bibfnamefont {Z.}~\bibnamefont {Wang}}, \bibinfo {author} {\bibfnamefont {Z.}~\bibnamefont {Xiang}}, \bibinfo {author} {\bibfnamefont {Z.}~\bibnamefont {Qiao}}, \bibinfo {author} {\bibfnamefont {J.}~\bibnamefont {Ying}},\ and\ \bibinfo {author} {\bibfnamefont {X.}~\bibnamefont {Chen}},\ }\bibfield  {title} {\bibinfo {title} {Pressure-induced dimensional crossover in a kagome superconductor},\ }\href {https://doi.org/10.1103/PhysRevLett.128.077001} {\bibfield  {journal} {\bibinfo  {journal} {Phys. Rev. Lett.}\ }\textbf {\bibinfo {volume} {128}},\ \bibinfo {pages}
  {077001} (\bibinfo {year} {2022})}\BibitemShut {NoStop}%
\bibitem [{\citenamefont {Tsirlin}\ \emph {et~al.}(2023)\citenamefont {Tsirlin}, \citenamefont {Ortiz}, \citenamefont {Dressel}, \citenamefont {Wilson}, \citenamefont {Winnerl},\ and\ \citenamefont {Uykur}}]{Tsirlin_PRB2023}%
  \BibitemOpen
  \bibfield  {author} {\bibinfo {author} {\bibfnamefont {A.~A.}\ \bibnamefont {Tsirlin}}, \bibinfo {author} {\bibfnamefont {B.~R.}\ \bibnamefont {Ortiz}}, \bibinfo {author} {\bibfnamefont {M.}~\bibnamefont {Dressel}}, \bibinfo {author} {\bibfnamefont {S.~D.}\ \bibnamefont {Wilson}}, \bibinfo {author} {\bibfnamefont {S.}~\bibnamefont {Winnerl}},\ and\ \bibinfo {author} {\bibfnamefont {E.}~\bibnamefont {Uykur}},\ }\bibfield  {title} {\bibinfo {title} {Effect of nonhydrostatic pressure on the superconducting kagome metal {CsV}$_{3}${Sb}$_{5}$},\ }\href {https://doi.org/10.1103/PhysRevB.107.174107} {\bibfield  {journal} {\bibinfo  {journal} {Phys. Rev. B}\ }\textbf {\bibinfo {volume} {107}},\ \bibinfo {pages} {174107} (\bibinfo {year} {2023})}\BibitemShut {NoStop}%
\bibitem [{\citenamefont {Koepernik}\ and\ \citenamefont {Eschrig}(1999)}]{Koepernik_PRB1999}%
  \BibitemOpen
  \bibfield  {author} {\bibinfo {author} {\bibfnamefont {K.}~\bibnamefont {Koepernik}}\ and\ \bibinfo {author} {\bibfnamefont {H.}~\bibnamefont {Eschrig}},\ }\bibfield  {title} {\bibinfo {title} {Full-potential nonorthogonal local-orbital minimum-basis band-structure scheme},\ }\href {https://doi.org/10.1103/PhysRevB.59.1743} {\bibfield  {journal} {\bibinfo  {journal} {Phys. Rev. B}\ }\textbf {\bibinfo {volume} {59}},\ \bibinfo {pages} {1743} (\bibinfo {year} {1999})}\BibitemShut {NoStop}%
\end{thebibliography}
%

%

\end{document}